\documentclass{sig-alternate}

\usepackage{amsmath,epsfig}
\usepackage{amsfonts}

\usepackage{PageSetting}
\usepackage{wrapfig, epsfig, algorithm, algorithmic, graphicx}

\def\done{\hspace*{\fill} $\framebox[1.5mm]{}$\smallskip}

\title{Can the Utility of Anonymized Data be used for Privacy Breaches?}

\numberofauthors{1}

\author{
  Raymond Chi-Wing Wong$^1$, Ada Wai-Chee Fu$^2$,
  Ke Wang$^3$, Yabo Xu$^3$, Philip S. Yu$^4$ \vspace*{0.4cm} \\
\begin{tabular}{c c c}
$^1$ Hong Kong University of Science and Technology & $^2$Chinese University of Hong Kong \\
\normalsize{raywong@cse.ust.hk} & \normalsize{adafu@cse.cuhk.edu.hk} \\
 $^3$ Simon Fraser University & $^4$ University of Illinois at Chicago\\
\normalsize{\{wangk,yxu\}@cs.sfu.ca} & \normalsize{psyu@cs.uic.edu}
\end{tabular}
}

\begin{document}

\begin{sloppy}

\maketitle

\begin{abstract}
Group based anonymization is the most widely studied approach for
privacy preserving data publishing. This includes $k$-anonymity, $l$-diversity,
and $t$-closeness, to name a few. The 
goal of this paper is to raise a fundamental
issue on the privacy exposure of the current group based approach. This has
been overlooked in the past. The group based anonymization approach basically
hides each individual record behind a group to preserve data privacy. If not
properly anonymized, patterns can actually be derived from the published data
and be used by the adversary to breach individual privacy. For example, from
the medical records released, if patterns such as people from certain countries
rarely suffer from some disease can be derived, then the information can be used
to imply linkage of other people in an anonymized
group with this disease
with higher likelihood. We call the derived patterns from the published data
the foreground knowledge. This is in contrast to the background knowledge that
the adversary may obtain from other channels as studied in some previous work.
Finally, we
show by experiments that the attack is realistic in the privacy benchmark
dataset under the traditional group based anonymization approach.
\end{abstract}

\category{H.2.0.a}{Information Technology and Systems}{Database Management}[General]

\terms{Algorithms, Experimentation, Security}


\section{Introduction}
\label{sec:intro}


A major technique used in privacy preservation data publishing is
\emph{group based anonymization}, whereby records in the given relation are
partitioned into groups and each group must ensure some property
such as diversity so as to satisfy the privacy requirement while
maintaining sufficient data utility. There are many privacy models
associated with group based anonymization such as $k$-anonymity
\cite{sweeney-kanonymity-model}, $l$-diversity \cite{l-diversity},
$t$-closeness \cite{LL07}, ($k, e$)-anonymity \cite{ZKS+07},
Injector \cite{LL08} and $m$-confidentiality \cite{WFW+07}.  It
\emph{seems} that this technique is sound for privacy preserving
data publishing. However, when examined more carefully, they all
suffer from one fundamental privacy violation problem, which is
overlooked in the past. The main cause of this problem is that the
\emph{utility} that is maintained in the anonymzied table can help
the adversary to breach individual privacy.

In the literature, background knowledge
\cite{l-diversity,KG06,MKM+07,WFW+07,LL08} such as the rarity of a
disease among a certain ethnic group or the pattern of age or gender
for a disease can be used by the adversary to breach individual
privacy.
 In this paper, we show that
  such knowledge can be \emph{mined} from the \emph{published data}
or the \emph{anonymized data} to compromise individual privacy. In
fact, one of the main purposes of data publishing is data mining
which
is mainly about the discovery of patterns from the published data. 

Let us illustrate the problem with an example. Suppose a table $T$
is to be anonymized for publication. Table $T$ has two kinds of
attributes, the quasi-identifier (QI) attributes and the sensitive
attribute. The QI attributes
 can be used as an identifier
in the table.
\cite{sweeney-kanonymity-model} points out that in a real dataset, most 
individuals can be uniquely identified by three QI attributes,
namely sex, date of birth and 5-digit zip code. The sensitive
attribute contains some sensitive values. In our example,
Table~\ref{tab:rawData} is the given table $T$ where one of the QI
attributes is ``Nationality" and the sensitive attribute is
``Disease" containing sensitive values such as Heart Disease and
HIV. Note that there can be other QI attributes in this table such
as sex and zip code. For the sake of illustration, we list
attribute ``Nationality" only.
 Assume that each tuple in the table is owned by an
individual and each individual owns at most one tuple.

Suppose that we want to anonymize $T$ and publish the anonymized
dataset $T^*$ to satisfy some privacy requirements.
Typically, $T^*$ consists of a set of \emph{anonymized groups} (in
short, \emph{A-groups}), 
where each A-group is a set of tuples with a multi-set of sensitive
values that are linked with the A-group. Depending on the
anonymization mechanism, each A-group may correspond to either a set
of quasi-identifer (QI) values or a single generalized QI value. An
attribute GID is added for the ID of the A-group. Such an anonymized
dataset is generated as a result of \emph{group-based anonymization}
commonly adopted in the literature of data publishing
\cite{AggarwalICDT05,Incognito,XT06b,WFW+07,LL08,LL07} (including
$k$-anonymity, $l$-diversity, $t$-closeness and a vast number of
other privacy models).

For illustration, a simplified setting of the $l$-diversity model
\cite{l-diversity} is used as a privacy requirement for published
data $T^*$. An A-group is said to be \emph{$l$-diverse} or satisfy
\textit{$l$-diversity} if in the A-group the number of occurrences
of any sensitive value is at most $1/l$ of the group size. A table
satisfies $l$-diversity (or it is
$l$-diverse) if all A-groups in it are $l$-diverse. 
Table~\ref{tab:genTable} satisfies 2-diversity. The intention is
that each individual cannot be linked to a disease with a
probability of more than 0.5. However, \emph{does this table protect
individual privacy sufficiently?}

\begin{table}
\center
\small
\begin{tabular}{ c | c | c | c |}\cline{2-4}
  Name & Nationality & ... & Disease \\ \hline
  Alex & American & ... & Heart Disease \\ \cline{2-4}
  Bob & Japanese & ... & Flu\\ \cline{2-4}
   & Japanese & ... & Flu \\ \cline{2-4}
   & Japanese &  ... & Stomach Virus\\ \cline{2-4}
   & French & ... & HIV\\ \cline{2-4}
   & Japanese & ... & Diabetes \\ \cline{2-4}
   & ... & ... & ... \\ \cline{2-4}
\end{tabular}
\caption{An example} \label{tab:rawData}
\vspace*{-0.4cm}
\end{table}

\begin{table}
\center
\small
\begin{tabular}{c c}
\begin{minipage}[htbp]{4cm}
\center
\begin{tabular}{| c | c | c |}\hline
  Nationality &  ... & GID\\ \hline
  American &  ... & $L_1$ \\ \hline
  Japanese &  ...  & $L_1$ \\ \hline
  Japanese &  ... & $L_2$ \\ \hline
  Japanese & ... & $L_2$ \\ \hline
  French &  ...  & $L_3$ \\ \hline
  Japanese & ... & $L_3$ \\ \hline
  ... & ... & ... \\ \hline
\end{tabular}
\end{minipage}
& \hspace*{0mm}
\begin{minipage}[htbp]{4cm}
\center
\begin{tabular}{| c | c |}\hline
  GID & Disease \\ \hline
  $L_1$ & Heart Disease \\ \hline
  $L_1$ & Flu\\ \hline
  $L_2$ & Flu \\ \hline
  $L_2$ & Stomach Virus\\ \hline
  $L_3$ & HIV\\ \hline
  $L_3$ & Diabetes\\ \hline
  ... & ...\\ \hline
\end{tabular}
\end{minipage}
\\
(a) QI Table &   (b) Sensitive table
\end{tabular}
\caption{A 2-diverse dataset anonymized from
Table~\ref{tab:rawData}} \label{tab:genTable} \label{tab4}
\vspace*{-0.4cm}
\end{table}

Let us examine the A-group with GID equal to $L_1$
as shown in Table~\ref{tab:genTable}. 
We also refer to the A-group by $L_i$. In $L_1$, Heart Disease and
Flu are values of the sensitive attribute Disease. It \emph{seems}
that each of the two individuals, Alex and Bob, in this group has a
50\% chance of linking to Heart Disease (Flu). The reason why the
chance is interpreted as 50\% is that the analysis is based on this
group \emph{locally} without any additional information.

However, from the \emph{entire published table} containing
\emph{multiple} groups, the adversary may discover some
interesting patterns \emph{globally}.
For example, suppose the published table consists of many A-groups
like $L_2$ with all Japanese with no occurrence of Heart Disease. At
the same time, there are many A-groups like $L_3$ containing some
Japanese without Heart Disease. The pattern that Japanese rarely
suffer from Heart Disease can be uncovered. Note that it is very
likely that such an anonymized data is published by conventional
anonymization methods, given the fact that Heart Disease occurs
rarely among Japanese. With the pattern uncovered, the adversary can
say that Bob, being a Japanese, has less chance of having Heart
Disease. S/he can deduce that Alex, being an American, has a higher
chance of having Heart Disease. The intended 50\% threshold is thus
violated.

\subsection{Foreground Knowledge Attack}

The anonymized data can be seen as an \emph{imprecise} or \emph{uncertain data}
\cite{BDJRV05,BDRV07}, and an adversary can uncover interesting
patterns since the published data must maintain high data utility
\cite{XT06b,ZKS+07,WFW+07}.
We call the uncovered patterns the
\emph{foreground knowledge} (which is \emph{implicitly}
inside the table) in contrast to the \emph{background knowledge},
studied by existing works
\cite{l-diversity,LL07,ZKS+07,WFW+07},
which the adversary requires much \emph{effort} to obtain
from somewhere outside the table.
Since it is easy to obtain the foreground knowledge
from the anonymized dataset, all existing works suffer from
privacy breaches.

In Table~\ref{tab:genTable}, 
there are only two \emph{local possible worlds} for assigning the
disease values to the two individuals in $L_1$: (1) $w_1:$ Alex is
linked to Heart Disease and Bob is linked to Flu and (2) $w_2:$ Alex
is linked to Flu and Bob is linked to Heart Disease. To construct a
probability distribution over the domain of the real world, a
simplest definition is based on the assumption that \emph{all the
possible worlds are equally likely}, or \emph{each world has the
same probability}.

If we publish a group $L_1$ alone, the random world assumption is a
good principle in the absence of other information. However, when
several groups are published together as typically the case, the
groups with Japanese contribute to a statement that their members are
not likely linked to Heart Disease. This statement means that the
\emph{probability} (or \emph{weight}) of the possible world $w_1$ is
much greater than that of $w_2$.


Most previous privacy works such as
$l$-diversity \cite{l-diversity},
$t$-closeness \cite{LL07}, ($k, e$)-anonymity \cite{ZKS+07} and
$m$-confidentiality \cite{WFW+07} adopt the random world assumption \emph{locally}.
In this paper, the source of attack of the adversary is to apply the
more complete model of the \emph{weighted possible worlds}.
We call this kind of attack \emph{foreground knowledge attack}.

\subsection{Contributions}

Our contributions can be summarized as follows.
%
  Firstly, 
  we define
  and study
  data anonymization issues in data publication with the consideration of
  foreground knowledge attack, which is ignored in the privacy literature.
  Secondly, 
we show how an adversary can breach privacy by computing the
probability that
  an individual is linked to a sensitive value by using foreground knowledge.

    Finally,
      we have conducted experiments
    to show how the adversary can succeed in foreground knowledge attack
    for four recent privacy models, namely \emph{Anatomy} \cite{XT06b}, \emph{MASK} \cite{WFW+07}, \emph{Injector} \cite{LL08}
and \emph{$t$-closeness} \cite{LL07}. 

We emphasize that, similar to $l$-diversity, \emph{all} privacy
models using group-based anonymization \cite{XT06b,WFW+07,LL08,LL07}
also suffer from possible \emph{privacy breaches} due to the
\emph{utility} of the published table. We believe that this work is
significant in pointing out this overlooked issue, and that all
followup works should need to deter foreground knowledge attack.

The rest of the paper is organized as follows.
Section~\ref{sec:probDef} formulates the problem.
Section~\ref{sec:algGroupPrivacy} describe how the adversary can
breach individual privacy with the foreground knowledge obtained
from the anonymized data. 
Section~\ref{alg:algForeground} shows how
the adversary can obtain the foreground knowledge from the
anonymized data. 
%
An empirical
study is reported in Section~\ref{sec:exp}.
Section~\ref{sec:related}  reviews the related work.
The paper is
concluded in Section~\ref{sec:concl}.

\section{Problem Definition}
\label{sec:probDef}

Let $T$ be a table. We assume that one of the attributes is a
sensitive attribute $X$ where some values of this attribute should
not be linkable to any individual. The value of the sensitive
attribute of a tuple $t$ is denoted by $t.X$. A
\textit{quasi-identifier} (QI) is a set of attributes of $T$,
namely $A_1, A_2, ..., A_q$, that may serve as identifiers for
some individuals. Each tuple in the table $T$ is related to one
individual and no two tuples are related to the same individual.

Let $P$ be a partition of table $T$.
We give a unique ID called GID to this partition $P$
and append an additional attribute called GID to this partition where
each tuple in $P$ has the same GID value.
Existing group-based anonymization defines a function $\beta$ on $P$ to
form an A-group such that
the linkage between the QI attributes
and the sensitive attribute in the A-group is lost.
There are two ways in the literature for this task.
One is \emph{generalization} by generalizing
all QI values to the same value. The other is
\emph{bucketization} by
forming two tables, called the \emph{QI
table} and the \emph{sensitive table}, where
$P$ is projected on all QI attributes and attribute GID
to form the QI table, and on
the sensitive attribute and attribute GID to form the
sensitive table.
 A table $T$ is \emph{anonymized} to a dataset $T^*$ if
 $T^*$ is formed by first partitioning $T$ into a number
 of partitions, then forming an A-group from
 each partition by $\beta$ and finally inserting each A-group into $T^*$.
%
For example, Table~\ref{tab:rawData} is anonymized to
Table~\ref{tab:genTable} by bucketization.


In known voter registration lists,
the QI values can often be used to identify a unique individual
\cite{sweeney-kanonymity-model,Incognito}. We assume that there is a
mapping which maps each tuple in $T$ to an A-group in $T^*$.
For example, the first tuple $t_1$ in
Table~\ref{tab:rawData} is mapped to A-group $L_1$.

In the following, for the sake of illustration,
we focus on discussing the anonymized table generated
by bucketization, instead of generalization.
The discussion for generalization is same as that for bucketization.
Specifically,
generalization is similar to bucketization but generalization
changes all QID values in a partition to the same ``generalized" values.
If the table is generated by generalization, each A-group contains the same ``generalized" values.
In the worst case scenario (which is a basic assumption in the privacy literature \cite{MKM+07,WFW+07,LL09}), the adversary
can uniquely map each individual in an A-group by an external table such as a voter registration
list. After the mapping, each A-group contains individuals with the original QID values, which
becomes the case of bucketization. Thus, the discussion for bucketization
still applies in the case for generalization.
The worst case scenario assumption is essential in  data publishing.
Nobody can afford if the privacy of an individual is breached \cite{LL09}.
AOL published the dataset about search logs in 2006.
After it realized that a \emph{single} 62 year old woman
living Georgia can be re-identified
from the search logs by New York Times reporters, it withdraws
the search logs and fired two employers responsible
for releasing the search logs \cite{BJ06}.

In the literature \cite{XT06b,WFW+07,LL08,LL07}, it is assumed that
the knowledge of the adversary includes (1) the published dataset
$T^*$, (2) the QI value of a
target individual, 
(3) an external table $T^e$ such as voter registration list that
maps QIs to individuals
\cite{sweeney-kanonymity-model,Incognito}.
We also follow these assumptions in our analysis.

The aim of privacy preserving data publishing is to deter any attack
from the adversary on linking an individual to a certain sensitive
value. Specifically, the data publisher would try to limit the probability that
such a linkage can be established. Let us consider an arbitrary
sensitive value $x$ for the analysis.
We denote any value in $X$ which is not $x$ by $\overline{x}$. 

In this paper, we consider that an adversary can obtain additional
information from the published dataset $T^*$ in the form of
\emph{global distribution}, which can lead to individual privacy
breach. In the example in Section~\ref{sec:intro}, we can mine from
the published table that the chance of Japanese suffering from Heart
Disease is low compared with American. This pattern is from the
\emph{global distribution} for the attribute set \{``Nationality"\}.

Consider an arbitrary sensitive value ``Heart Disease".
Table~\ref{tab:globalDistMotivatingExample} shows the \emph{global
distribution} of attribute set \{``Nationality"\}, which consists of
the probabilities that a Japanese, an American or a French is linked
to Heart Disease. Each probability in the table is called a
\emph{global probability}. The \emph{sample space} for each such
probability consists of the possible assignments of the values $x$
and $\overline{x}$ to an individual with the particular nationality.

Each possible value in attribute ``Nationality" is called a
\emph{signature}. There are three possible signatures in our
example: ``Japanese", ``American" and ``French". In general, there
are other attribute sets, such as \{``Sex", ``Nationality"\}, with
their correspondence global distributions. We define the signature
and the global distribution for a particular attribute set
$\mathcal{A}$ as follows.

\begin{table}
\center
\small
\begin{tabular}{| c | c | c |} \hline
  $p()$ & Heart Disease & Not Heart Disease \\ \hline
  American & 0.1 & 0.9 \\
  Japanese & 0.003 & 0.997 \\
 French & 0.05 & 0.95 \\ \hline
\end{tabular}
\caption{A global distribution of attribute ``Nationality"
for our motivating example} \vspace*{-0.4cm}
\label{tab:globalDistMotivatingExample}
\end{table}

\begin{definition}[Signature]
Let $T^*$ be the published dataset. Given a QI attribute set
$\mathcal{A}$ with $r$ attributes $A_1, ..., A_r$. A
\emph{signature} $s$ of $\mathcal{A}$ is a set of attribute-value
pairs $(A_1, v_1),...,(A_r, v_r)$ which appear in the published
dataset $T^*$, where $A_i$ is a QI attribute and $v_i$ is a value. A
tuple $t$ in $T^*$ is said to \emph{match} $s$ if $t.A_i=v_i$ for
all $i = 1, 2, ..., r$.
\end{definition}

For example, a signature $s$ can be \{(``Nationality", ``American"),
(``Sex", ``Male")\} if the attribute set $\mathcal{A}$
is \{``Nationality", ``Sex"\}. 
For convenience, we often drop the attribute names in a signature,
and thus we refer to \{``American", ``Male"\} instead of
\{(``Nationality", ``American"), (``Sex", ``Male")\}. The first
tuple $t_1$ in Table~\ref{tab:genTable}(a) matches \{``American"\}
but the second tuple does not.

\begin{definition}[Global Distribution]
Given an attribute set $\mathcal{A}$, the \emph{global distribution}
$G$ of $\mathcal{A}$ contains a set of entries $(s:x, p)$ for each
possible signature $s$ of $\mathcal{A}$, where $p$ is equal to
$p(s:x)$ which denotes the probability that a tuple matching
signature $s$ is linked to $x$ given the published dataset $T^*$.
\end{definition}

For example, if $G$ contains (``Japanese":``Heart Disease", 0.003)
and (``American":``Heart Disease", 0.1), then the probability that a
Japanese patient is linked to Heart Disease is equal to 0.003 while
that of an American patient is 0.1.

%

The global distribution $G$ derived from the published dataset $T^*$
is called the \emph{foreground knowledge}. We will describe how the
adversary derives $G$ from the published
table. 

\begin{problem}[Foreground Knowledge]
Given any arbitrary attribute
   set $\mathcal{A}$, we want to find the global distribution $G$ of
   $\mathcal{A}$ from published dataset $T^*$.
\label{problem:foregroundKnowledge}
\end{problem}

From Section~\ref{sec:intro}, we show that with the global
distribution $G$ of attribute set \{``Nationality"\}, we can deduce
that the chance of Alex, an American, suffering from Heart Disease
is high.
Let $t$ be Alex and $x$ be Heart Disease. The chance can be
formulated by $p(t:x)$, the probability that $t$ is linked to $x$
given $G$.
%

\begin{problem}[Privacy Breach]
Given a published dataset $T^*$, for any individual $t$, any
sensitive value $x$ and any attribute set $\mathcal{A}$, we want to
determine whether the probability that $t$ is linked to $x$ denoted
by $p(t:x)$ is greater than $1/r$. Individual $t$ is said to suffer
from privacy breaches if the probability is greater than $1/r$.
\label{problem:individualPrivacyBreach} 
\end{problem}


In this paper, we first study Problems~\ref{problem:foregroundKnowledge}
and \ref{problem:individualPrivacyBreach}. In
Section~\ref{sec:algGroupPrivacy}, we will first give how we solve
Problem~\ref{problem:individualPrivacyBreach} assuming
that we are given 
the foreground knowledge. Then, in Section~\ref{alg:algForeground},
we will describe how we can mine the foreground knowledge from the
published dataset $T^*$ for
Problem~\ref{problem:foregroundKnowledge}. We shall show that the
two problems are intertwined, since the global probability is
derived based on the published table, and thus depends on the
probability $p(t:x)$ for each tuple $t$.

\begin{table*}[tbp]
\center
\begin{tabular}{|c|c|c|}
  \hline
   $ p()$ & $x$ &  ${\overline{x}}$\\ \hline
  $s_1$ & $f_{1}$ & $\overline{f_{1}}$ \\
  $s_2$ & $f_{2}$ & $\overline{f_{2}}$ \\
  : & :  & : \\
  \hline
\end{tabular}
\caption{Global distribution}
\label{tab:globalDistributionTableGeneral}
\end{table*}

\begin{table*}
\center 
\begin{tabular}{c c}
\begin{minipage}[ht]{3cm}
\begin{center}
\begin{tabular}{ | c | c |c |}\hline
  $p()$ & $x$ &  $\overline{x}$ \\ \hline
  $s_1$ & 0.5 & 0.5 \\ 
  $s_2$ & 0.2 & 0.8 \\ \hline
\end{tabular}
\end{center}
\end{minipage}
\\
(a) Global distribution
\\
\begin{minipage}[ht]{11cm}
\begin{center}
\begin{tabular}{|c| c | c | c | c | c | c |} \hline
 $w$ & $t_1 $ & $t_2$ & $t_3$ & $t_4$ & $p(w)$ & $p(w|L_k)$\\
  & $(s_1)$&$(s_1)$&$(s_2)$& $(s_2)$ & & \\ \hline
$w_1$ &  $x$ & $x$ & $\overline{x}$ & $\overline{x}$ &  $0.5 \times
0.5 \times 0.8 \times 0.8 = 0.16$ &  $0.16/0.33 = 0.48$  \\ \hline
 $w_2$ & $x$ & $\overline{x}$ & $x$ & $\overline{x}$ &  $0.5 \times 0.5 \times 0.2 \times 0.8 = 0.04$ &  $0.04/0.33 = 0.12$ \\ \hline
$w_3$ &  $x$ & $\overline{x}$ & $\overline{x}$ & $x$ &  $0.5 \times
0.5 \times 0.8 \times 0.2 = 0.04$ &  $0.04/0.33 = 0.12$\\ \hline
  $w_4$ & $\overline{x}$ & $x$ & $x$ & $\overline{x}$ &  $0.5 \times 0.5
\times 0.2 \times 0.8 = 0.04$ &  $0.04/0.33 = 0.12$\\ \hline
 $w_5$ & $\overline{x}$ & $x$ & $\overline{x}$ & $x$ &  $0.5 \times 0.5 \times 0.8 \times 0.2 = 0.04$ &  $0.04/0.33 = 0.12$ \\ \hline
  $w_6$ &$\overline{x}$ & $\overline{x}$ & $x$ & $x$ &  $0.5 \times 0.5 \times 0.2 \times 0.2 = 0.01$ &  $0.01/0.33 = 0.03$ \\ \hline
\end{tabular}
\end{center}
\end{minipage}
\\
 (b) $p(w)$ and $p(w|L_k)$
\end{tabular}
\caption{An example illustrating the computation of $p(t_j:x)$}
\label{tab:exampleIllustrateProbLocalComputation} 
\end{table*}

\begin{table}[htbp]
\center \small
\begin{tabular}{|c|l|}
  \hline
  $L_k$ & an A-group (anonymized group) in the \\ & anonymized dataset\\
  $\mathcal{A}$ & set of attributes
  e.g. \{``Nationality", ``Sex"\} \\
  $t_1,...,t_N$ & tuples in an $A$-group \\
  $s_1,...,s_m$ & signatures for $\mathcal{A}$, e.g.\{``American", ``Male"\}\\
  & multiple tuples $t_j$'s can map to the
  same $s_i$ \\
  $x$ & a sensitive value \\
  $\bar{x}$ & any value not equal to $x$\\
  $p(t_j:x)$ & probability that tuple $t_j$ is linked to
  value $x$ \\
  $p(s_i:x)$ & probability that signature $s_i$ is linked to $x$ \\
   $f_i$ &  a simplified notation for $p(s_i:x)$ \\
  $\bar{f_i}$ & $1 - f_i$ \\
  $w$ & a possible world: an assignment of the tuples\\
  & in A-group $L_k$ to the
  sensitive values $x$ and $\bar{x}$ \\
$\mathcal{W}_k$ & set of all possible worlds $w$ for $L_k$\\
$B$ &  set of all possible worlds $w$ in $\mathcal{W}_k$\\
&
in which $t_j$ is assigned value $x$.\\
  $p(w)$ & probability that $w$ occurs given the anonymized \\
  & dataset and based on $\mathcal{A}$ \\
  $p(w|L_k)$ & conditional probability that $w$ occurs given \\ & A-group $L_k$
  \\
  $p_{j,w}$ & let $t_j$ be linked to $\gamma$ in $w$, where $\gamma$ is $x$ or $\overline{x}$
  \\ & $p_{j,w}$ is
  the probability that $t_j$ is linked to $\gamma$\\
 $\mathcal{L}_{s_i}$ &  set of A-groups containing
tuples matching $s_i$
\\
$L_k(s_i)$ & the set of tuples in $L_k$ matching $s_i$.\\
$c_k(s_i:x)$ & the expected number of tuples which match $s_i$
\\ &
and are linked to $x$ in the A-group $L_k$\\
  \hline
  \end{tabular}
  \caption{Notations}
\label{tab:notation} \vspace*{-0.3cm}
\end{table}

\section{Finding Privacy Breaches}
\label{sec:algGroupPrivacy}

We assume that the attack is based on the linkage of an attribute set
$\mathcal{A}$ to a sensitive value $x$. We denote by $\bar{x}$ any
value not equal to $x$. In this section, we assume that the global
distributions $G$ for $\mathcal{A}$ and $x$ have been determined and
we show how an adversary can use $G$ to find privacy breaches.
How the global distributions can be derived is explained in
Section~\ref{alg:algForeground}.

Suppose there are $m$ possible signatures for attribute set
$\mathcal{A}$, namely $s_1, s_2, ..., s_m$. The global distribution
$G$ of $\mathcal{A}$ is shown in
Table~\ref{tab:globalDistributionTableGeneral}. To simplify our
presentation, the probability that $s_i$ is linked to $x$
($\overline{x}$),
$p(s_i:x )$ ($p(s_i:\overline{x} )$), is denoted by $f_i$
($\overline{f}_i$).


Given $G$, the formula for $p(t:x)$, the probability that a tuple
$t$ is linked to sensitive value $x$, is derived here. Suppose $t$ belongs to
A-group $L_k$. For the ease of reference, let us summarize the notations
that we use in Table \ref{tab:notation}. We shall need the following
definitions.

\begin{definition}[primitive events, projected events]
A mapping $t:\gamma$ from an individual or tuple $t$ to a sensitive
value $\gamma$ ($x$ or $\bar{x}$) is called a \emph{primitive event}.
Suppose $t$ matches signature $s$. Let us call an event for the
corresponding signature, ``$s:\gamma$", a \emph{projected event} for
$t$.
\end{definition}

Hence, a primitive event is an event in the sample space for
$p(t:x)$, which is the probability of the interest for the
adversary. A projected event is an event for $p(s:x)$ which appears
in the global distribution $G$.

\begin{definition}[possible world]
Consider an A-group $L_k$ with $N$ tuples, namely $t_1, t_2, ...,
t_N$, with sensitive values $\gamma_{1}, \gamma_{2}, ... \gamma_{N}$, where $\gamma_i$
is either $x$ or $\overline{x}$ for $i = 1, 2, ..., N$. A possible
world $w$ for $L_k$ is a possible assignment mapping the tuples in set $\{t_1,
t_2, ..., t_N\}$ to values in multi-set $\{\gamma_{1}, \gamma_{2}, ...
\gamma_{N}\}$ in $L_k$.
\end{definition}

Given an A-group $L_k$ with a set of tuples and a multi-set of
sensitive values. For each possible world $w$, according to the
global distribution $G$ based on attribute set $\mathcal{A}$, we
compute the probability $p(w)$ that $w$ occurs. The sample space for
$p(w)$ consists of all the possible assignments of $x$ or
$\overline{x}$ to a set of $N$ tuples with the same signatures as
those in $L_k$.

Suppose that in a possible world $w$ for $L_k$, tuple $t_j$ is
linked to $\gamma$, where $\gamma$ is either $x$ or $\overline{x}$.
Let $p_{j, w}$ be the probability that $t_j$ is
linked to $\gamma$. 


Like \cite{l-diversity,XT06b,WFW+07}, we assume that the linkage of a sensitive value to an individual is
independent of the linkage of a sensitive value to another
individual. For example, whether an American suffers from Heart
Disease is independent of whether a Japanese suffers from Heart
Disease. Thus, for a possible world $w$ for $L_k$, the probability
that $w$ occurs is the product of the probabilities of the
corresponding projected events for the tuples $t_1, ... t_N$ in
$L_k$.
\begin{eqnarray}
  p(w ) =  p_{1, w} \times p_{2, w} \times ... \times p_{N, w}
  \label{eqn1}
\end{eqnarray}

Suppose $t_j$ matches signature $s_i$.
If $t_j$ is linked to $x$ in $w$, then $p_{j, w} = f_{i}$.
Otherwise, $p_{j, w} = \overline{f_{i}}$.

$p(w)$ corresponds to the \emph{weight} of $w$, which we mentioned
in the introduction.

The probability of $L_k$ given $T^*$ is the sum of the probabilities
of all the possible worlds consistent with $T^*$ for $L_k$. Let the
set of these worlds be $\mathcal{W}_k$. For $w \in \mathcal{W}_k$,
we have
\begin{eqnarray}
p(w|L_k) = \frac{ p(w)}{\mbox{$\sum_{w' \in \mathcal{W}_k}$}
p(w')} \label{eqn2}
\end{eqnarray}

It is easy to verify that 
$\sum_{w \in \mathcal{W}_k} p(w|L_k) = 1$. \\

Our objective is to find the probability that an individual $t_j$
 in $L_k$ is linked to a sensitive value $x$.
 This is given by the sum of the conditional probabilities
 $p(w |L_k)$ of all the possible
 worlds $w$ where $t_j$ is linked to $x$.
\begin{eqnarray}
  p(t_j:x) = \mbox{$\sum_{w \in B_x}$} p(w | L_k)
   \label{eqn-tupleLinkage}
\end{eqnarray}
where $B_x$ is a set of all possible 
worlds $w$ in
$\mathcal{W}_k$ in which $t_j$ is assigned value $x$.

One can verify that 
 $p(t_j:x ) +  p(t_j:\overline{x} ) = 1$.

\begin{example} 
Consider an A-group $L_k$ in a published table $T^*$. Suppose there
are four tuples, $t_1, t_2, t_3$ and $t_4$, and four sensitive
values, $x, x, \overline{x}$ and $\overline{x}$ in $L_k$. Suppose
the published table $T^*$ satisfies 2-diversity.

Consider the global distribution $G$ based on a certain QI
attribute set $\mathcal{A}$ which contains two possible signatures
$s_1$ and $s_2$ as shown in
Table~\ref{tab:exampleIllustrateProbLocalComputation}(a).

Suppose $t_1, t_2, t_3$ and $t_4$ match signatures $s_1, s_1, s_2$
and $s_2$, respectively. There are six possible worlds $w$ as shown
in Table~\ref{tab:exampleIllustrateProbLocalComputation}(b). For
example, the first row is the possible world $w_1$ with mapping
\{$t_1:x$, $t_2:x$, $t_3:\overline{x}$, $t_4:\overline{x}$\}. The
table also shows
the probability $p(w)$ of the possible worlds. Take the first
possible world $w_1$ for illustration. From the global distribution
in Table~\ref{tab:exampleIllustrateProbLocalComputation}(a),
$p(s_1:x) = 0.5$ and $p(s_2:\overline{x}) = 0.8$. Hence,
$p(w_1) = 0.5 \times 0.5 \times 0.8 \times 0.8 = 0.16$.
The sum of probabilities
$p(w)$ of all possible worlds from
Table~\ref{tab:exampleIllustrateProbLocalComputation}(b) is equal to
0.16 + 0.04 + 0.04 + 0.04 + 0.04 + 0.01 = 0.33. Consider
 $w_1$ again. Since $p(w_1) = 0.16$,
$p(w_1|L_k) = 0.16/0.33 = 0.48$.

Suppose the adversary is interested in the probability that $t_1$ is
linked to $x$. We obtain $p(t_1 : x )$ as follows. $w_1, w_2$ and
$w_3$, as shown in
Table~\ref{tab:exampleIllustrateProbLocalComputation}(b), contain
``$t_1:x$".
Thus, $p(t_1:x )$ is equal to the sum of the probabilities
$p(w_1|L_k), p(w_2|L_k)$ and $p(w_3|L_k)$.
$ p(t_1 : x ) 
= 0.48 + 0.12 + 0.12  = 0.72 $ which is greater than 0.5, the
intended upper bound for 2-diversity that an individual is linked to
a sensitive value.\done
\end{example}

Let $|L_k|$ be the size of the A-group containing $t_j$ and
$|\mathcal{W}_k|$ be the number of possible worlds in an A-group
$L_k$.
We will generate $|\mathcal{W}_k|$ possible worlds.
For each possible world, we calculate $p(w)$ and $p(w|L_k)$
in $O(|L_k|)$ time.
Thus, the time complexity is $O(|L_k| \cdot
|\mathcal{W}_k|)$. 

The time complexity depends
on two factors. One is
$|\mathcal{W}_k|$ and another is $|L_k|$.
(1) $|\mathcal{W}_k|$ is equal to $C^{N}_{n}$ where $n$ is the number of tuples with $x$ in this A-group of size $N$ and $C^{N}_{n}$ denotes the total number of possible ways of choosing $n$ objects from $N$ objects.
Note that $\mathcal{W}_k$ is typically small because $n$ is usually equal to a small
number. For $l$-diversity, algorithm Anatomy \cite{XT06b} restricts that each A-group contains either $l$ or $l+1$ tuples and each sensitive value $x$ appears at most once. Here, $n$ is equal to 1. Thus, for each possible $x$, $|\mathcal{W}_k|$ is at most $l+1$. For Algorithm MASK \cite{WFW+07}, in our experiment with $l = 2$, the greatest frequency of $x$ in an A-group is 8. The size of this A-group is 23. $|\mathcal{W}_k|$ is equal to $C^{23}_8 = 490,314$. When $l=10$, the greatest possible value of $|\mathcal{W}_k|$ is 140,364,532. These values are small compared with the excessive number of possible worlds studied in uncertain data \cite{IJ84,BDJRV05,BDRV07,AKO07,CCM+08} (e.g., $10^{10^6}$ in \cite{AKO07})). In the experimental setups in existing works \cite{l-diversity,XT06b,LL07,WFW+07,LL08}, $l$ $\leq 10$. In other words, $\mathcal{W}_k$ can be generated within a reasonable time.
(2) $|L_k|$ is bounded by the greatest size of the A-group which depends on
the anonymization techniques. For example, $|L_k|$ is equal to $l$ or $l+1$ for
algorithm Anatomy \cite{XT06b} restricting that each A-group contains either $l$ or $l+1$.
In our experiment, $|L_k|$ is at most 23 for algorithm MASK \cite{WFW+07} where $l = 2$.

\section{Mining Foreground Knowledge}
\label{alg:algForeground}

We first describe how we find the global distribution $G$ of a
certain attribute set $\mathcal{A}$ from the anonymized data in
Section~\ref{subsec:global}. Next, we introduce a pruning strategy to
prune our search space of attribute sets in Section~\ref{subsec:foregroundMultipleAttributeSet}.
Finally, we describe the algorithm for
finding the global distribution of multiple attribute sets and
discuss its complexity in Section~\ref{subsubsec:foregroundMiningAlg}.

\subsection{Foreground Knowledge}
\label{subsec:global}

In the previous section, we assume that the values of $f_i$ are
given. Here we consider how to derive $f_{i}$ from the published
table $T^*$. We will develop $m$ equations involving the $m$
variables $f_i$, 
$1 \leq i \leq
m$. 

 Let the set of A-groups in $T^*$ be
$L_1, ..., L_u$. 
Let $L_k(s_i)$ be the set of tuples in $L_k$ matching signature
$s_i$. For example, in Table~\ref{tab:genTable}, let $s_i =
$\{``American"\}. Then, $L_1(s_i)$ contains only the first tuple.


Let $\mathcal{L}_{s_i}$ be a set of A-groups containing tuples which
match $s_i$. That is, $\mathcal{L}_{s_i} = \{ L_k | L_k(s_i) \neq
\emptyset \}$.

 $f_{i}$ is equal to the expected number of
tuples which match $s_i$ and are linked to $x$ in $T^*$ divided by
the number of tuples which match $s_i$ in $T^*$. Let $c_k(s_i:x)$ be
the expected number of tuples which match $s_i$ and are linked to
$x$ in the A-group $L_k$. Then, we can express $f_{i}$ as follows.
\begin{eqnarray}
\label{eqn3}
 f_{i} = \frac{{\mbox{$\sum_{L_k \in {\mathcal{L}_{s_i}}}$}
c_k(s_i:x)}} {\mbox{$\sum_{L_k \in \mathcal{L}_{s_i}}$} |L_k(s_i)|}
\end{eqnarray}
The denominator is simply equal to the number of occurrences of $s_i$
in $T^*$ and which can be easily found from the dataset. Let us
consider the term $c_k(s_i:x)$ in the numerator.

Without additional knowledge to govern otherwise, we assume that the
event that a tuple matching $s_i$ in $L_k$ is linked to $x$ is
independent of the event that another tuple also matching $s_i$ in
$L_k$ is linked to $x$. Then we have the following.
 \begin{eqnarray}
c_k(s_i:x) = |L_k(s_i)| \times p(t_j:x)
 \label{eqn4}
\end{eqnarray}
where $t_j$ is any tuple in $L_k$ matching $s_i$. Note that any
$t_j$ in $L_k$ matching $s_i$ can be used here since all such
$p(t_j:x)$ values are equal.
%
Substitute Equations (\ref{eqn-tupleLinkage}) and (\ref{eqn2}) into the above
equation, we get
\begin{eqnarray}
c_k(s_i:x) = |L_k(s_i)| \times \mbox{$\sum_{w \in B_x}$} \frac{
p(w)}{\mbox{$\sum_{w' \in \mathcal{W}_k}$} p(w')} \label{eqn5}
\end{eqnarray}
Hence, $c_k(s_i:x)$ is expressed in terms of probabilities $p(w)$
 which in turn are expressed in the $m$
variables $f_i$ (see Equation (\ref{eqn1})
where $p_{j, w}$ is equal to $f_i$ or $\overline{f}_i$). Here
note that $\overline{f_i} = 1 - f_i$.

There are $m$ equations of the form of Equation~(\ref{eqn3}) for the
expression of $f_i$, $1 \leq i \leq m$. These equations involve $m$
variables, $f_i$. This is a classical problem of a system of
simultaneous non-linear equations, which occurs in many
applications. It can be solved by conventional methods such as
Newton's method and Bairstow's iteration.
Since Newton's method \cite{CC02} has been known to be effective and
feasible, we choose this method for our study in
this paper. 
%

\label{sec:newtonExample}

\begin{example}  
Given a table $T$ containing six tuples, $t_1, t_2, ..., t_6$, as shown in
Table~\ref{tab:exampleIllustratingForegroundAccuracy_orgTable}.
If the objective of the privacy requirement is 2-diversity, $T$ does
not satisfy 2-diversity. Thus, an anonymized dataset $T^*$
Table~\ref{tab:exampleIllustratingGlobal} with three A-groups, $L_1,
L_2$ and $L_3$, is published (for each sensitive value $x$ and each
A-group, the fraction of tuples with $x$ is at most 0.5).
Note that $L_3$ satisfies 2-diversity because,
Since $\overline{x}$
corresponds to a value not equal to $x$,
in $L_3$, the first $\overline{x}$
corresponds to a value $y$
and the second $\overline{x}$ corresponds to another value $z$.
%

\begin{table}[tbp]
\center
\small
\begin{tabular}{ c c c c}
\begin{minipage}[ht]{2.0cm}
\center
\begin{tabular}{| c | c | c |}\hline
  $\mathcal{A}$ & ... & $X$ \\ \hline
  $s_1$ & ...& $x$ \\ \hline
  $s_1$ & ...& $x$ \\ \hline
  $s_1$ & ...& $\overline{x}$ \\ \hline
  $s_2$ & ...& $\overline{x}$ \\ \hline
  $s_2$ & ...& $\overline{x}$ \\ \hline
  $s_2$ & ...&  $\overline{x}$ \\ \hline
\end{tabular}
\caption{A raw table}
\label{tab:exampleIllustratingForegroundAccuracy_orgTable}
\end{minipage}
&
\begin{minipage}[ht]{6.0cm}
\noindent
\begin{tabular}{c c}
\begin{minipage}[htbp]{2.5cm}
\center
\begin{tabular}{c | c | c | c |}\cline{2-4}
  $t$ & $\mathcal{A}$ & ... & GID \\ \hline
  $t_1$ & $s_1$ & ... & $L_1$ \\ \cline{2-4}
  $t_2$ & $s_2$ & ... & $L_1$ \\ \cline{2-4}
  $t_3$ & $s_1$ & ... & $L_2$ \\ \cline{2-4}
  $t_4$ & $s_1$ & ... & $L_2$ \\ \cline{2-4}
  $t_5$ & $s_2$ & ... & $L_3$ \\ \cline{2-4}
  $t_6$ & $s_2$ & ... & $L_3$ \\ \cline{2-4}
\end{tabular}
\end{minipage}
&
\begin{minipage}[htbp]{2.6cm}
\center
\begin{tabular}{| c | c |}\hline
  GID & 
        {$X$} \\ \hline
  $L_1$ & $x$ \\ \hline
  $L_1$ & $\overline{x}$ \\ \hline
  $L_2$ & $x$ \\ \hline
  $L_2$ & $\overline{x}$\\ \hline
  $L_3$ & $\overline{x}$\\ \hline
  $L_3$ & $\overline{x}$\\ \hline
\end{tabular}
\end{minipage}
\\ & \\
(a) QI Table & (b) Sensitive Table
\end{tabular}
\caption{An example illustrating the computation of the global distribution}
\label{tab:exampleIllustratingGlobal} \vspace*{-0.5cm}
\end{minipage}
\end{tabular}
\end{table}

Consider the global distribution of attribute set $\mathcal{A}$.
There are two possible signatures based on $\mathcal{A}$, namely
$s_1$ and $s_2$. 
Thus, we have two equations with two variables, namely $f_{1}$
and $f_{2}$, the probabilities
in the global distribution $G$ of $\mathcal{A}$ as shown in
Table~\ref{tab:globalDistributionTableGeneral}.

Consider $f_{1}$. Since only A-groups $L_1$ and $L_2$ contain the
tuples matching $s_1$, $\mathcal{L}_{s_1} = \{L_1, L_2\}$.
$$
f_{1} = [\mbox{$\sum_{L_k \in \mathcal{L}_{s_1}}$} c_k(s_1:x)]/[\mbox{$\sum_{L_k \in \mathcal{L}_{s_1}}$} |L_k(s_1)|] 
$$
$L_1$ contains one tuple $t_1$ matching $s_1$ and $L_2$ contains two
tuples $t_3, t_4$ matching $s_1$, $|L_1(s_1)| = 1$ and $|L_2(s_1)| =
2$. Thus,
\begin{equation}
f_{1} = [1 \times p(t_1:x) + 2 \times p(t_3:x)]/(1 + 2) 
\label{eqn:f11}
\end{equation}
Consider $L_1$. There are only two possible worlds, $w_1 = \{t_1:x,
t_2:\overline{x}\}$ and $w_2 = \{t_1:\overline{x}, t_2:x\}$.
Note that $t_1$ and $t_2$ match signatures $s_1$ and $s_2$, respectively.
$p_{1, w_1} = f_1, p_{2, w_1} = \overline{f}_2,
p_{1, w_2} = \overline{f}_1$ and $p_{2, w_2} = f_2$.
Thus,
$p(w_1) = p_{1, w_1} \times p_{2, w_1} = f_{1} \times \overline{f_{2}}$ and $p(w_2) =
p_{1, w_2} \times p_{2, w_2} = \overline{f_{1}} \times f_{2}$. We derive that
$$
p(t_1:x) = p(w_1 | L_1) = f_{1}\overline{f_{2}}/(f_{1}\overline{f_{2}} + \overline{f_{1}}f_{2}) 
$$
Similarly, consider $L_2$. There are two possible worlds, $w_3 =
\{t_3:x, t_4:\overline{x}\}$ and $w_4 = \{t_3:\overline{x},
t_4:x\}$. Similarly, $p(w_3) = f_{1}\times \overline{f_{1}}$ and
$p(w_4) = \overline{f_{1}} \times f_{1}$. We have
$$
p(t_3:x) = p(w_4|L_2) = f_{1}\overline{f_{1}}/(f_{1}\overline{f_{1}} + \overline{f_{1}}f_{1}) = 1/2 
$$
From (\ref{eqn:f11}), we obtain
\begin{eqnarray*}
 f_{1} & = & [f_{1}\overline{f_{2}}/(f_{1}\overline{f_{2}} + \overline{f_{1}}f_{2}) +
 1]/3 \\
 & = & [ f_1 ( 1- f_2) / (f_1 (1-f_2) + (1-f_1)f_2) + 1]/3
\end{eqnarray*}

Similarly, since $L_1$ contains one tuple $t_2$ matching $s_2$ and
$L_3$ contains two tuples $t_5, t_6$ matching $s_2$,
\begin{eqnarray*}
f_{2}
  & = & [1 \times p(t_2:x) + 2 \times p(t_5:x)]/(1+2)  \\
  & = & [\overline{f_{1}}f_{2}/(f_{1}\overline{f_{2}} + \overline{f_{1}}f_{2}) + 0]/3 \\
& = &  [ (1 - f_1) f_2 / ( f_1 (1-f_2) + (1-f_1)f_2)]/3
\end{eqnarray*}
With the above two equations involving two variables, we adopt
Newton's method to solve for these variables. 

Finally, we obtain $f_{1} = 0.666667$ and $f_{2} = 0.000000$. Thus,
we derive  $\overline{f_{1}} = 0.333333$ and $\overline{f_{2}} =
1.000000$. 
\label{example:exampleIllustrateGlobal} \done
\end{example}

\subsection{Pruning Attribute Sets}
\label{subsec:foregroundMultipleAttributeSet}

The adversary may choose to attack with as many attribute sets as
possible. Although there are many attribute sets in the anonymized
data, it is not always true that the global distribution of each
attribute set is \emph{reliable} because if the global distribution
derived is based on a small sample or a small set of tuples matching
the same signature, the distribution is not accurate. For example,
consider attribute set $\mathcal{A}$=``Nationality" and the
signature \{``American"\}. Suppose there are only a few Americans,
says 10 Americans, in the published table $T^*$. Intuitively, 10
Americans cannot represent a meaningful global distribution. We will
make use of the sample size studied in the literature of statistics
to determine whether the distribution is reliable or not. The
adversary can launch an attack only based on reliable distributions.

Based on studies in statistics \cite{Toivonen96}, we use the
following theorem to determine the acceptable sample size (i.e., the
size of the set which contains the tuples matching the same signature $s$). Let $S$ be a random sample of tuples for a
signature $s$,
and $p$ be the expected fraction of tuples in $S$ with the
sensitive value $x$. Let $\widetilde{p}$ be the observed fraction of
tuples with the sensitive value $x$ in the sample $S$. Then the
following theorem applies.


\begin{theorem}[Sample Size \cite{Toivonen96}]
Given an error parameter $\epsilon \ge 0$ and a confidence parameter
$\sigma \ge 0$, if random sample $S$ has size $
  |S| \ge \frac{1}{2\epsilon^2} \ln \frac{2}{\sigma}
$, the probability that $|\widetilde{p} - p| > \epsilon$ is at most
$\sigma$. \label{thm:sampleSize} \done
\end{theorem}

In case the sample size is not enough to satisfy the error bound,
then uniform distribution will be assumed. The sample size satisfies
the monotonicity property. Formally, without loss of generality,
assume that there are $u$ attributes, namely $A_1, ..., A_u$. Let
$v_1 \in A_1, ..., v_u \in A_u$. Let $y(v_1, ..., v_{i})$ be the
number of tuples with attributes $(A_1, ..., A_i)$ equal to $(v_1,
..., v_i)$.
%
%
Given a positive integer $J$, if $y(v_1, ..., v_{i}) < J$, then
$y(v_1, ..., v_{i}, v_{i+1}) < J$.
%
With the above monotonicity property, whenever we find that the
sample size of $y(v_1, ..., v_{i})$ is not large enough, we do not
need to count the number of the tuples with values $v_1, ...,
v_{i+1}$ because $y(v_1, ..., v_{i}, v_{i+1})$ is also not large
enough. Thus, this can help to prune the search
space.

\subsection{Algorithm} \label{subsubsec:foregroundMiningAlg}

In this section, we will describe how to compute the set $\mathcal{G}$ of all
global distributions of multiple attribute sets with the use of
the sample size just described. 
The steps are shown in Algorithm \ref{alg1}.

\begin{algorithm}
\caption{Computation of the global distributions} \label{alg1}
\begin{itemize}
\item
\emph{Step 1:} For each attribute set $\mathcal{A}$, we first
identify the set $\mathcal{S}_{\mathcal{A}}$ of signatures $s_i$
with respect to $\mathcal{A}$ where each $s_i$ is matched by some
tuples in $T^*$ and has sufficient sample size. For example, for
$\mathcal{A}=\{$``Nationality", ``Sex" \}, a signature equal to \{
``American", ``Male"\} is matched by the first tuple in
Table~\ref{tab4}(a). If it has sufficient sample size, it is stored
in $\mathcal{S}_{\mathcal{A}}$.
\item
\emph{Step 2:} For each attribute set $\mathcal{A}$, if
$\mathcal{S}_{\mathcal{A}}$ is non-empty, we calculate the global
distribution of $\mathcal{A}$ according to
$\mathcal{S}_{\mathcal{A}}$ for each sensitive value $x$.
\end{itemize}
\end{algorithm}

In the algorithm, Step 1 is to find all signatures with sufficient
sample size for each attribute set $\mathcal{A}$. Similar to
frequent pattern mining, this step is typically computed within a
reasonable time.
Let $\alpha$ be the time for this step. 
After we have determined the sample sizes, $\mathcal{G}$ is used to
store the global distributions of all attribute sets each of which
contains signatures with sufficient sample size.

Step 2 is to calculate the global distribution of $\mathcal{A}$
according to non-empty $\mathcal{S}_{\mathcal{A}}$ for each
attribute set $\mathcal{A}$. In other words, it is to find each
global distribution in $\mathcal{G}$. As described in
Section~\ref{subsec:global}, for a particular global distribution,
we formulate $m$ equations with $m$ variables where $m$ is the total
number of signatures for $A$. The average number of terms in each
equation is $O(N \cdot |\mathcal{W}_k| \cdot |\mathcal{L}_{s_i}|)$
where $N$ is the average A-group size, $|\mathcal{W}_k|$ is the
average number of possible worlds in an A-group $L_k$ and
$|\mathcal{L}_{s_i}|$ is the average number of A-groups with tuples
matching a signature $s_i$. If Newton's method takes $\beta$ time to
find a solution, the computation for a global distribution takes
$O(m \cdot N \cdot |\mathcal{W}_k| \cdot |\mathcal{L}_{s_i}| +
\beta)$ time. Since there are $|\mathcal{G}|$ global distributions,
Step 2 takes $O(|\mathcal{G}| \cdot (m \cdot N \cdot |\mathcal{W}_k|
\cdot |\mathcal{L}_{s_i}| + \beta))$ time.
%

Thus, the total running time is
$O(\alpha + |\mathcal{G}| \cdot (m \cdot N \cdot |\mathcal{W}_k| \cdot
|\mathcal{L}_{s_i}| + \beta))$. Note that the values of $m$, $N$,
$|\mathcal{W}_k|$ and $ |\mathcal{L}_{s_i}|$ are small and the
complexity is dominated by $|\mathcal{G}|$ and $\beta$. But, as the attribute
set size increases, the sample size quickly becomes insufficient,
and so $|\mathcal{G}|$ is typically well-behaved.

From our experiments, 
in all of our cases, Step 2 with the system of $m$ equations can be
solved in a relatively
 short time. So, $\beta$ is also a
reasonable value. For the benchmark dataset, adult, foreground knowledge
can be mined within 12 minutes in all our experiments.

The probabilistic analysis is similar in nature to that studied for
uncertain databases 
\cite{BDJRV05,BDRV07,AKO07}
The
computation complexity above is in fact much smaller than these
previous works. 
In
\cite{AKO07}, all results are returned within 3 hours.
The reason is that \cite{BDJRV05,BDRV07,AKO07} analyze the possible worlds
based on the entire uncertain table (which can be regarded
as a single large A-group) while we analyze the possible
worlds based on a single small A-group (which is typically
smaller than the entire table).

\subsection{Discussion}

%

We have just discussed how to find the global distribution
from the published table. One may argue that the global distribution $\widetilde{\mathcal{G}}$
found from the published table is just an \emph{approximation} of
the \emph{true} global distribution $\mathcal{G}_o$ found from the original table.
Thus, the privacy breaches found in Section~\ref{sec:algGroupPrivacy}
according to $\widetilde{\mathcal{G}}$
are invalid. However, we disagree with this argument
with the following reasons.

Firstly, since the adversary does not have the true global
distribution $\mathcal{G}_o$ (because s/he has not seen
the original table), the best adversary's knowledge about the global
distribution is $\widetilde{\mathcal{G}}$.

Secondly, an adversary with $\widetilde{\mathcal{G}}$ is more \emph{powerful}
and more \emph{sophisticated} than another adversary without any knowledge
about the global distribution. The former adversary is what we are studying
in this paper and can breach individual privacy discussed in
Section~\ref{sec:algGroupPrivacy}
while the latter adversary is the normal adversary studied in the privacy
literature \cite{l-diversity,XT06b,WFW+07} and cannot breach any individual
privacy found by the former adversary.

Thirdly,
the adversary $\mathbb{A}_o$
with $\mathcal{G}_o$ (if there is) does not perform more
serious privacy attacks compared with
an adversary $\widetilde{\mathbb{A}}$ with $\widetilde{\mathcal{G}}$.
We assume that an adversary $\mathbb{A}_o$ can have the
true global distribution $\mathcal{G}_o$. This means that
the public can also know $\mathcal{G}_o$ and
$\mathcal{G}_o$ is not secret information\footnote{
If this is not true, one of the ways that adversary $\mathbb{A}_o$
can obtain $\mathcal{G}_o$ is to steal the original table from
the data publisher. Since s/he has the original table, the privacy
breaches found by $\mathbb{A}_o$ are more serious. In this paper,
we are not studying that the adversary can steal the original table.}.

Consider adversary $\widetilde{\mathbb{A}}$.
Before s/he obtains $\widetilde{\mathcal{G}}$,
individual privacy (in the published table) is protected.
After s/he obtains $\widetilde{\mathcal{G}}$ (which can be found
from the published table), individual privacy breaches.
There is a change of belief after s/he sees $\widetilde{\mathcal{G}}$.
There are two kinds of privacy breaches. 
The first one is that an adversary can guess correctly
the true sensitive value of an individual.
The second one is that s/he can guess incorrectly
the true sensitive value of an individual.
For example, even if an individual is not linked to HIV
in the original table, s/he can guess that the probability
that this individual is linked to HIV is very high. This
is also considered as a privacy breach to this individual.
The reason is that the disclosure of the high linkage
between this individual and HIV hurts the reputation of
the individual
because
the adversary can
convince a certain set of people that the inference procedure
about individual privacy breaches
was reasonable.
Thus, privacy breaches found by $\widetilde{\mathbb{A}}$
are also serious.

Consider adversary $\mathbb{A}_o$.
In this case, we know that $\mathcal{G}_o$ is a public information.
Thus, the data publisher must have already taken
$\mathcal{G}_o$ into the account to publish a table.
The claim is true because, otherwise, no individuals
are eager to disclose their information to data publisher.
Thus, even if adversary $\mathbb{A}_o$ sees $\mathcal{G}_o$,
we cannot breach any individual privacy in the published table.



\section{Empirical Study}
\label{sec:exp}

A Pentium IV 2.2GHz PC with 1GB RAM was used to conduct our
experiment. The algorithm was implemented in C/C++. We adopted the
publicly available dataset, Adult Database, from the UCIrvine
Machine Learning Repository \cite{UCIrvine}. This dataset (5.5MB)
was also adopted by
\cite{Incognito,l-diversity,WangKe-bootom-up,Wang-Top-down,WFW+07}.
We used a configuration similar to
\cite{Incognito,l-diversity,WFW+07}. The records with unknown values
were first eliminated resulting in a dataset with 45,222 tuples
(5.4MB). Nine attributes were chosen in our experiment, namely Age,
Work Class, Marital Status, Occupation,
Race, Sex, Native Country, Salary Class and Education. 
By default, we chose the first five 
attributes and the last
attribute 
as the
quasi-identifer and the sensitive attribute, respectively.
Similar to \cite{WFW+07}, in attribute ``Education",
all values representing the education levels before ``secondary" (or
``9th-10th") such as ``1st-4th", ``5th-6th" and ``7th-8th" are
regarded as a sensitive value set where an adversary checks whether
each individual is linked to this set more than $1/r$, where $r$ is
a parameter.

There are 3.46\% tuples with education levels before ``secondary".
We set $\epsilon = 0.01$ and $\sigma = 0.9$ for sampling. That is,
the allowed relative error of sampling is 1/3.46 = 28.90\%, which is
considered large. A larger allowed error means less attribute sets
can be pruned. Since there is a set $\mathcal{G}$ of multiple global
distributions $G$, we can calculate $p(t:x)$ for different $G$'s and
different $x$'s. We take the greatest such value to report as the
probability that individual $t$ is linked to some sensitive value
since this corresponds to the worst case privacy breach.

\subsection{Privacy Breach in $l$-diverse Tables}

In this section, we will show that foreground knowledge attack is
successful in the published data generated from the benchmark
dataset, adult, by a well-known privacy algorithm, \emph{Anatomy}
\cite{XT06b}. We set $l = r$ where $l$ is the parameter of
$l$-diversity used in Anatomy. We implemented the formula in
Section~\ref{sec:algGroupPrivacy} to calculate the probability of a
privacy breach and the formula in Section~\ref{alg:algForeground} to
find the global distribution from the published data.
If a tuple which appears in the published data is identified
as a privacy breach by our algorithm, it is said to be a \emph{problematic
tuple}.
The tuples linking
to sensitive values in the original table are called \emph{sensitive tuples}.
In this case study, we evaluate privacy breaches with three
measurements:
\begin{enumerate}\item \emph{proportion of problematic tuples among sensitive
tuples, (this is the recall in IR research).}
\item \emph{proportion of non-sensitive tuples which are
identified wrongly as problematic tuples by our algorithm},
 \item \emph{the average probability by which individual privacy is breached among
all sensitive tuples}.
\end{enumerate}

 We have conducted experiments with the variation of $r$
and the variation of the QI size. (1) Variation of $r$:
When $r = 2$ with default settings, the \emph{average probability that individual privacy breaches}
among all sensitive tuples is 0.8917($>1/2$). When $r$ is increased to
4, it becomes 0.4640($>1/4$).
%
When $r$ increases, there is a higher chance that
a tuple forms an A-group with other tuples. Thus,
the average size of A-groups is larger. Thus,
the average probability of privacy breaches decreases.
We also studied the \emph{proportion of problematic tuples}
among all sensitive tuples and the \emph{proportion of non-sensitive
tuples identified wrongly as privacy breaches}.

We found that, in most cases, more than 99\% of sensitive tuples
have privacy breaches and less than 6\% of non-sensitive tuples are
identified wrongly.
(2) Variation of the QI size:
When the QI size is equal to 3 with default settings where $r = 2$,
the average probability causing privacy breaches is 0.80307. When
the size is increased to 8, it becomes 0.943526. This is because
when there are more QI attributes, it is more likely that a QI
attribute (or attribute set) gives a global distribution which can
lead to privacy breaches.

We also have a case study in the published data
generated by \emph{Anatomy}. Suppose the QI attributes
chosen are Age, Marital Status and
Occupation and the sensitive attribute
is Education. In the original data,
there are the following 2 tuples.

\medskip

{ 
\begin{tabular}{| c | c | c | c |} \hline
  Age & Marital Status & Occupation & Education \\ \hline
  39 & Never-married & Adm-clerical & Bachelors \\ \hline
  20 & Married-civ-spouse & Craft-repair & 5th-6th \\ \hline
\end{tabular}
}

\medskip

Suppose the objective of \emph{Anatomy} is 2-diversity.
Since ``5th-6th" is a sensitive value, \emph{Anatomy} forms an
A-group containing these two tuples. However, from
the global distribution derived from the published
data with respect to attribute Occupation,
the probability that an individual with
Occupation=``Adm-clerical" is linked to a low education
is only 0.02 but the probability that an individual
with Occupation=``Craft-repair" is linked to a low
education is 0.04. Since there is a significant
difference in global distribution of attribute Occupation, the probability that
the second tuple above is linked to a low education
is 0.67 (which is greater than 0.5).


It is noted that the global distribution derived
from the published data matches the real situation
that ``Adm-clerical" jobs  require
higher educations but ``Craft-repair" jobs
does not. In other words, the foreground knowledge can
help the adversary to breach individual privacy.
%
More specifically, let us check whether the \emph{real} global
distribution derived from the original table
is similar to the global distribution derived
from the published data. From the original
table, the probability that an individual with
Occupation=``Adm-clerical" is linked to a low education
is only 0.01 but the probability that an individual
with Occupation=``Craft-repair" is linked to a low
education is 0.04. We observe that this global distribution
is similar to that derived from the published data.

With our default experimental setting using sufficient sample size,
for 2-diversity, the average relative error of the global
probabilities derived from the published data=0.7\% which achieves
99.3\% accuracy. For 10-diversity, the error increases to 5.26\%
where the accuracy is 94.74\%. It shows that statistically the
accuracy is very high. In other words, the foreground knowledge
derived from the published data is quite accurate compared with the
knowledge derived from the original table.


In all our experiments, privacy breaches can be found within 12
minutes, which shows that foreground knowledge attack can easily be
realized.

\subsection{Privacy Breach in Other Privacy Models}
\label{subsec:analysisProposed}

We studied privacy breaches with four algorithms,
\emph{Anatomy} \cite{XT06b}, \emph{MASK} \cite{WFW+07}, \emph{Injector} \cite{LL08}
and \emph{$t$-closeness} \cite{LL07}.
They are selected because they consider $l$-diversity or similar privacy
requirements, so we need
only set $l = r$.
For \emph{Anatomy}, we set $l = r$.
For \emph{MASK}, the parameters $k$ and $m$ used in MASK are set to
$r$. For \emph{Injector}, the parameters $minConf$, $minExp$ and $l$
are set to 1, 0.9 and $r$, respectively, which are
the default settings in \cite{LL08}. For \emph{$t$-closeness},
similar to \cite{LL07}, we set $t = 0.2$.
We evaluate the algorithms in terms of four measurements:
(1) \emph{time for mining foreground knowledge},
(2) \emph{execution time}, 
(3) the \emph{proportion of problematic tuples among all sensitive tuples},
(4) the average of the greatest difference in the global
probabilities in each A-group (In our figures, we label this as
\emph{average value of $\triangle$}), and (5) the \emph{relative error
ratio} in answering an aggregate query as in
\cite{XT06b,WFW+07,LL08} by the published data.
For each measurement, we
conducted the experiments 100 times and took the average.

We do not report the time for finding privacy breaches because the
time is very short (within a few minutes). For the sake of space,
since the proportion of non-sensitive tuples identified wrongly for
privacy breaches is small (less than 10\%), we do not report here.

Let us explain measurements (4) and (5). (4) Consider an A-group $L_k$ contains
two tuples matching signatures $s_i$ and $s_j$, respectively.
Suppose $p(s_i:x)$ is the greatest global probabilities and
$p(s_j:x)$ is the smallest in the A-group. The value of $\triangle$
in $L_k$ is equal to $p(s_i:x) - p(s_j:x)$. The average value of
$\triangle$ is taken among all A-groups and all attribute sets
$\mathcal{A}$ with sufficient samples. (5) The relative error ratio
measures the utility of the published data. We adopt all query
parameters in 
\cite{XT06b,WFW+07,LL08}.
For each evaluation, 
we performed 10,000 queries and reported the average relative
error ratio.

We have conducted the experiments by varying two factors: (1)
 the QI size, and (2) $r$. 


%

\begin{figure}[tb] 
\center
\begin{tabular}{c c}
    \begin{minipage}[htbp]{4.0cm}
        \includegraphics[width=4.0cm,height=3.0cm]{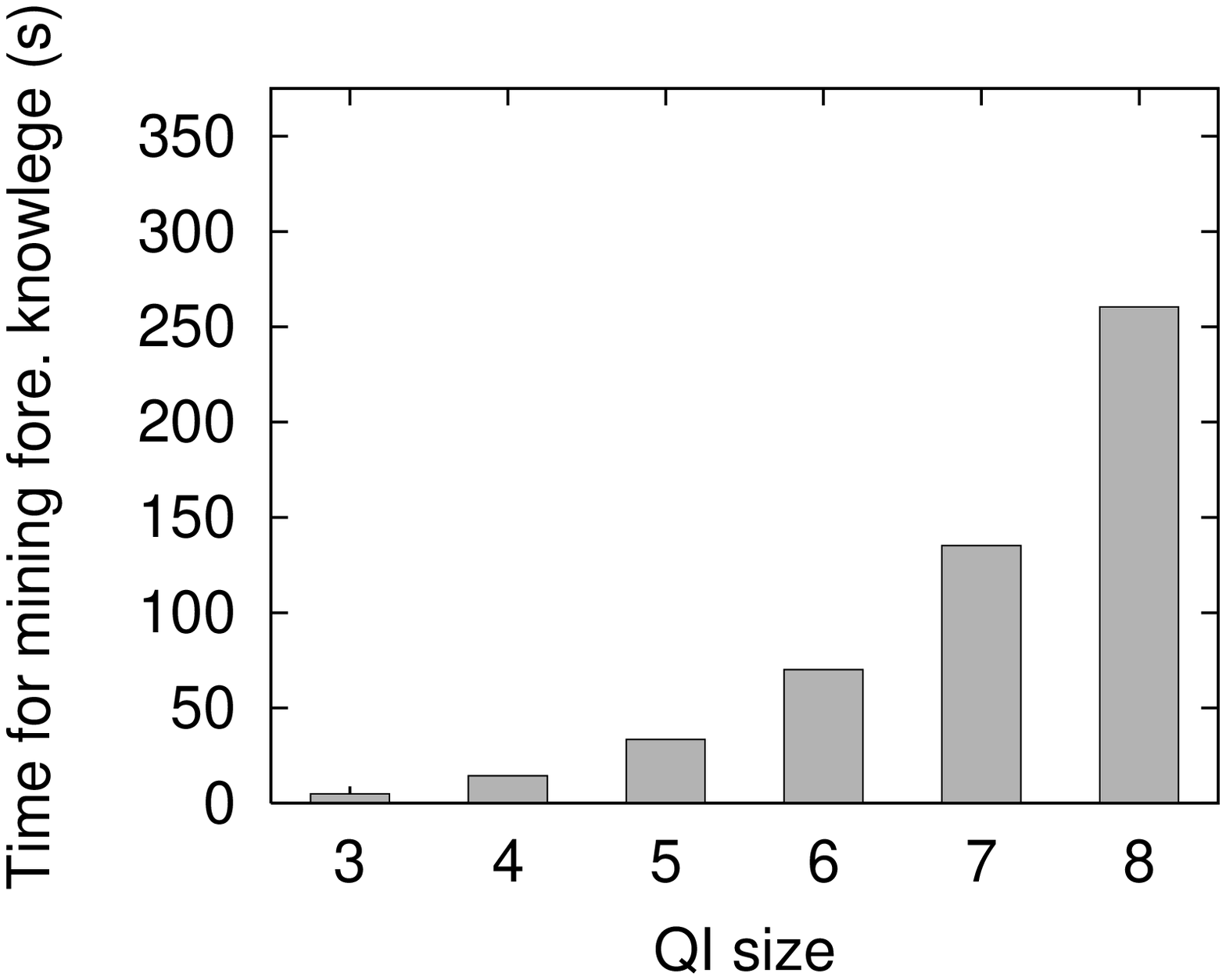}
    \end{minipage}
&
    \begin{minipage}[htbp]{4.0cm}
        \includegraphics[width=4.0cm,height=3.0cm]{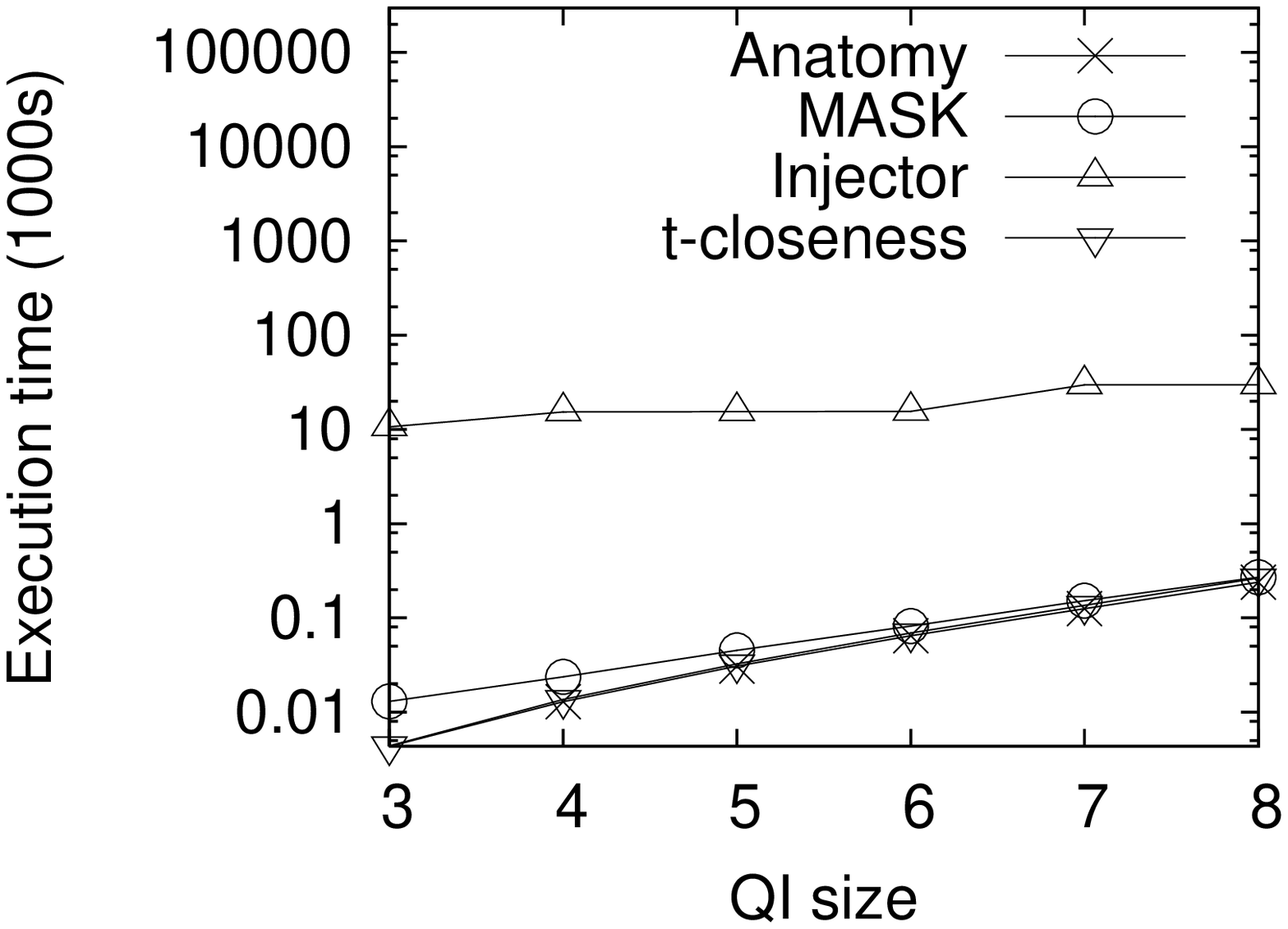}
    \end{minipage}
\\
(a)
&
(b)
\\
    \begin{minipage}[htbp]{4.0cm}
        \includegraphics[width=4.0cm,height=3.0cm]{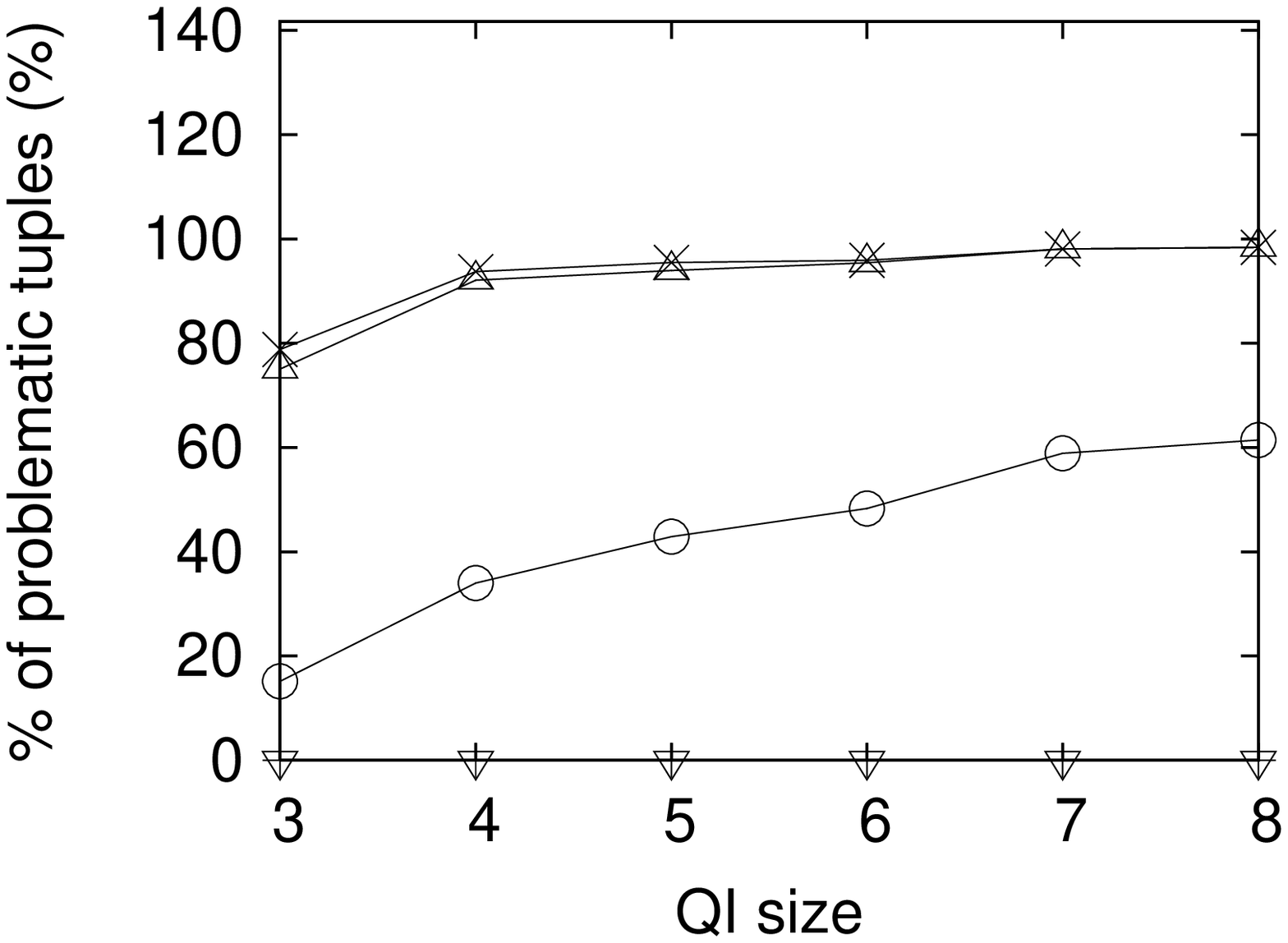}
    \end{minipage}
&
    \begin{minipage}[htbp]{4.0cm}
        \includegraphics[width=4.0cm,height=3.0cm]{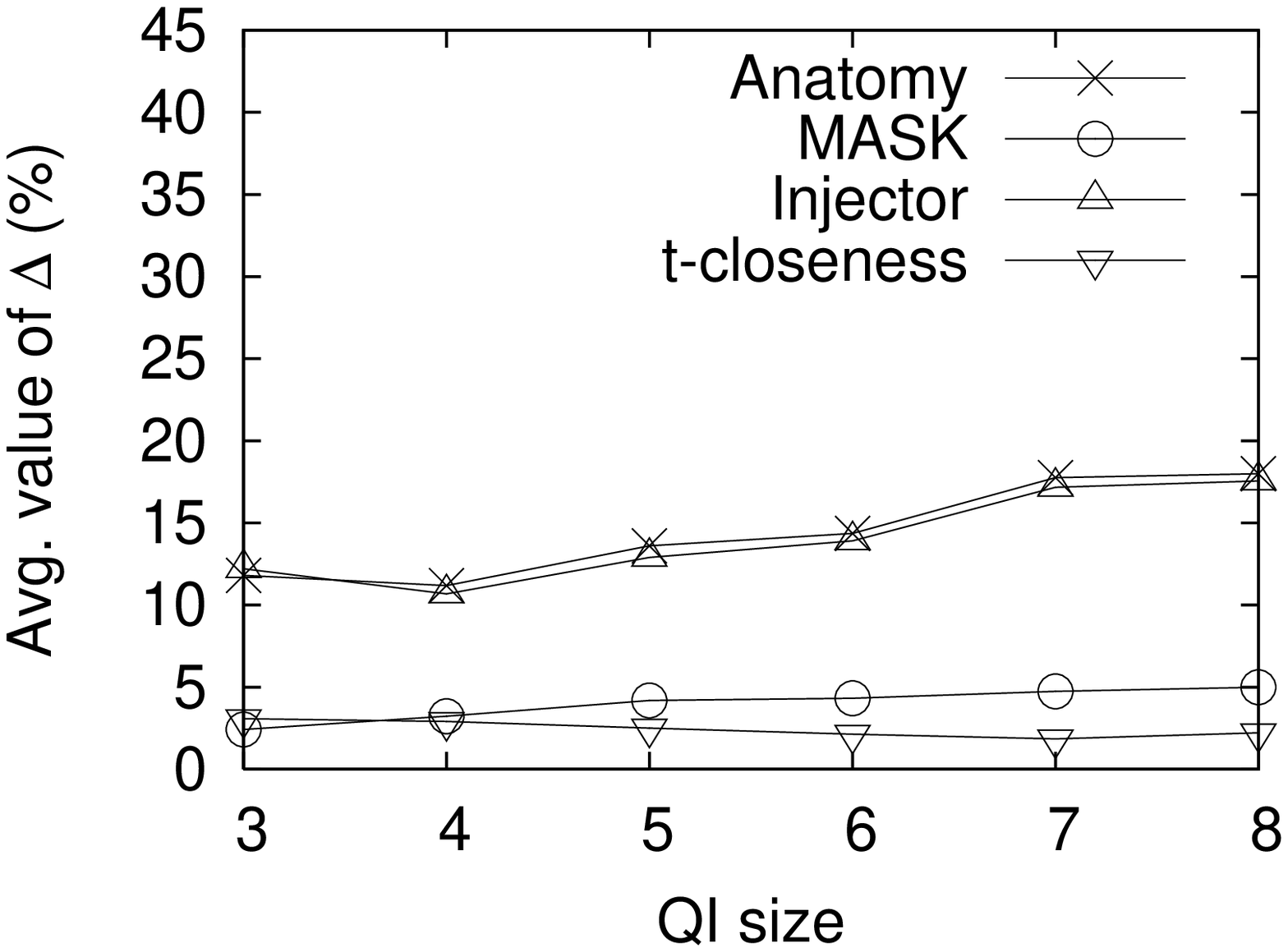}
    \end{minipage}
    \\
(c)
&
(d)
\end{tabular}
\caption{Effect of QI size ($r = 2$)}\label{fig:graphAgainstQID-m2}
\end{figure}

%

\begin{figure}[tb] 
\center
\begin{tabular}{c c}
    \begin{minipage}[htbp]{4.0cm}
        \includegraphics[width=4.0cm,height=3.0cm]{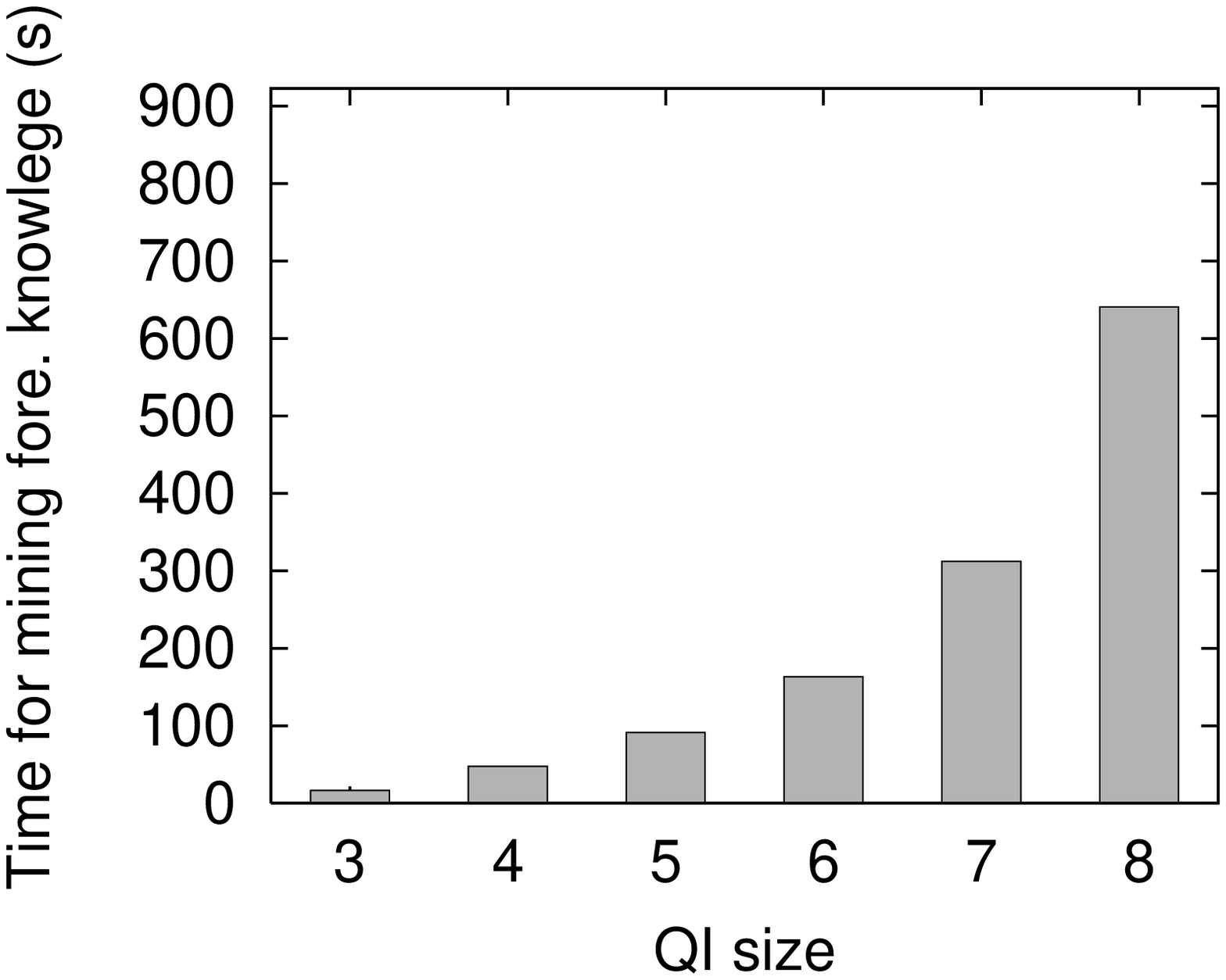}
    \end{minipage}
&
    \begin{minipage}[htbp]{4.0cm}
        \includegraphics[width=4.0cm,height=3.0cm]{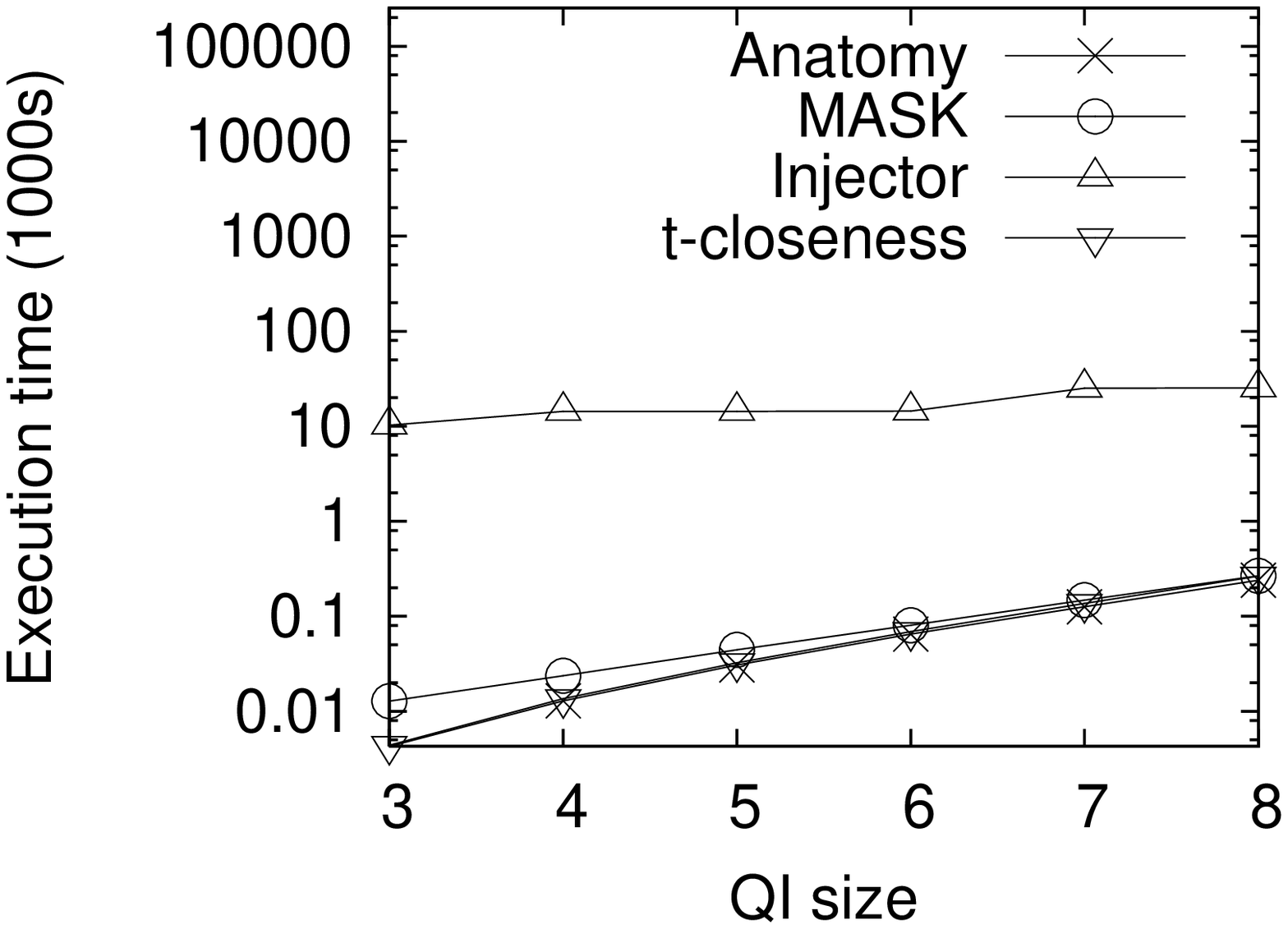}
    \end{minipage}
\\
(a)
&
(b)
\\
    \begin{minipage}[htbp]{4.0cm}
        \includegraphics[width=4.0cm,height=3.0cm]{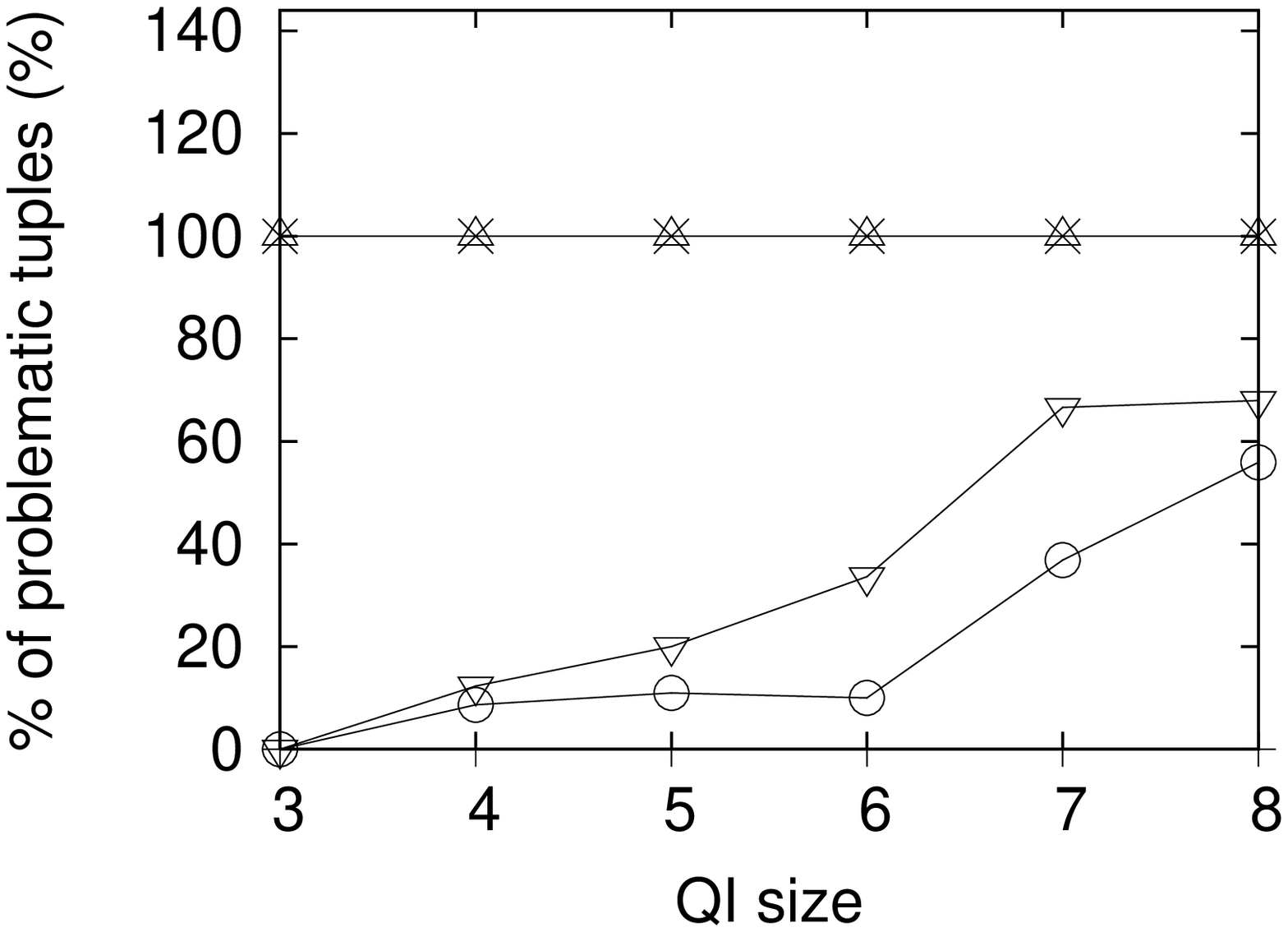}
    \end{minipage}
&
    \begin{minipage}[htbp]{4.0cm}
        \includegraphics[width=4.0cm,height=3.0cm]{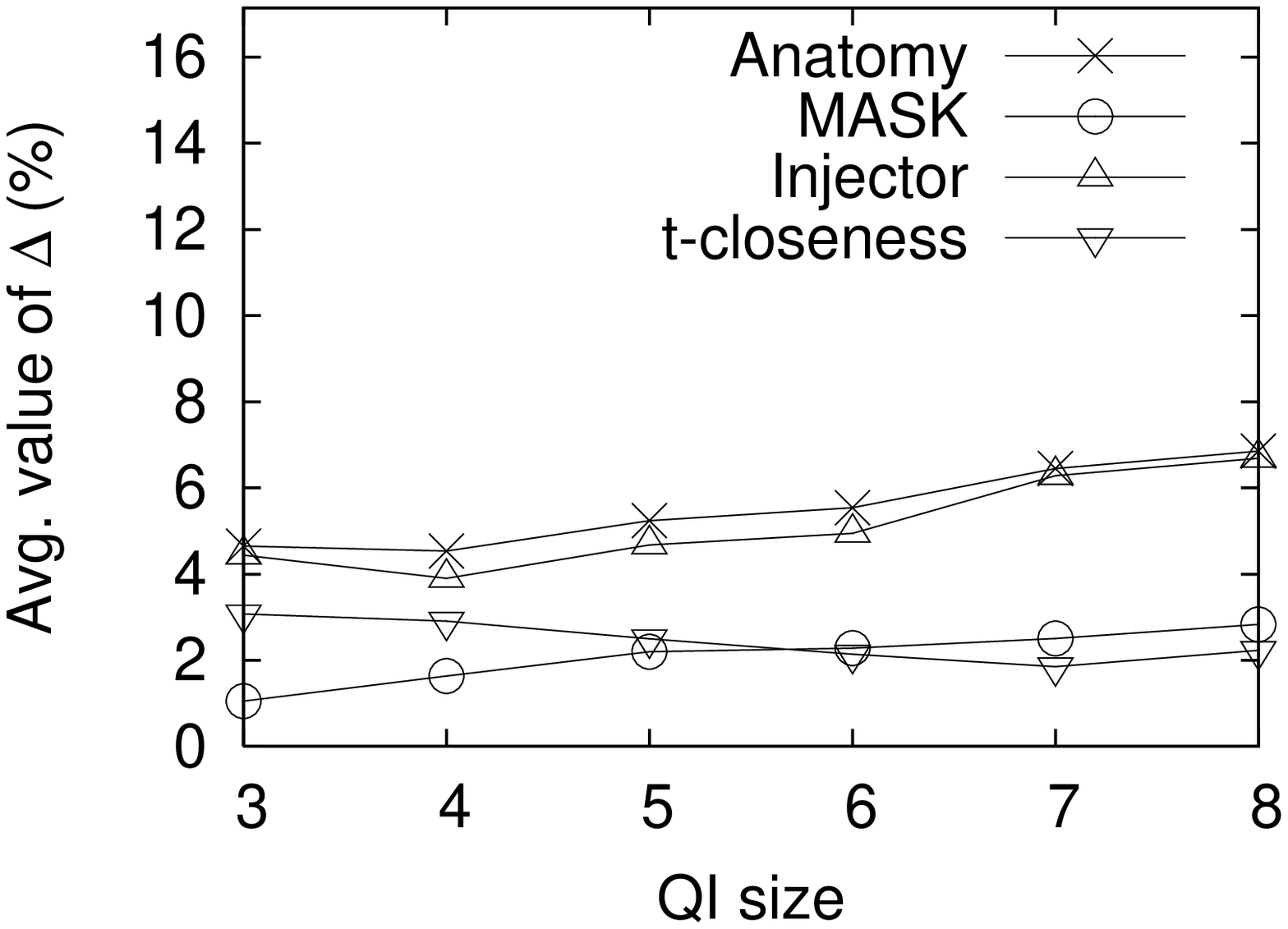}
    \end{minipage}
    \\
(c)
&
(d)
\end{tabular}
\caption{Effect of QI size ($r = 10$)}\label{fig:graphAgainstQID-m10}
\end{figure}

Figure~\ref{fig:graphAgainstQID-m2} and
Figure~\ref{fig:graphAgainstQID-m10} show the
results when $r$ is
set to 2 and 10, respectively.
Figure~\ref{fig:graphAgainstQID-m2}(a) shows that the time
for mining foreground knowledge increases with the QI size
because the algorithm needs to process more attribute sets.
Figure~\ref{fig:graphAgainstQID-m2}(b) shows that
the execution time increases with the QI size
because the algorithms have to process more
QI attributes.

Figure~\ref{fig:graphAgainstQID-m2}(c) shows that the proportion
of problematic tuples among sensitive tuples increases with QI size.
With a larger QI size, there is a higher chance that
individual privacy breaches due to more attributes which
can be used to construct the global distributions.
\emph{MASK} has fewer privacy breaches compared with
\emph{Anatomy} and \emph{Injector} because the side-effect of the minimization
of QI values in each A-group adopted in \emph{MASK}
makes the difference in the global distribution among
all tuples in each A-group smaller. Thus, the number of
individual with privacy breaches is smaller.
It is noted that there is no violation in 
\emph{$t$-closeness}. The reason why $t$-closeness
has no privacy breaches is due to the large A-groups formed
by global recoding with respect to value $r(=2)$.
The average size of the A-group in the table satisfying $t$-closeness
is at least 4000 and the utility of the table is low.
It is noted that parameter $t$ is independent of parameter $r$.
We will show that $t$-closeness has privacy breaches when $r = 10$.

In Figure~\ref{fig:graphAgainstQID-m2}(d),
when the QI size increases, the average value of $\triangle$ with respect to every attribute set increases, as shown in
Figure~\ref{fig:graphAgainstQID-m2}(d).
The average value of $\triangle$ is the largest in \emph{Anatomy} and \emph{Injector}, and
the third largest in \emph{MASK}. This is because
\emph{Anatomy} and \emph{Injector} does not take the global distribution directly into
the consideration for merging but \emph{MASK} does indirectly
during the minimization of QI values.

Figure~\ref{fig:graphAgainstQID-AvgRelativeError}(a) shows that
the average relative error of $t$-closeness is the largest since
it forms large A-groups by global recoding which introduce
a lot of errors and thus reduce the utility of the published data.

\begin{figure}[tbp] 
\center
\begin{tabular}{c c}
    \begin{minipage}[htbp]{4.0cm}
        \includegraphics[width=4.0cm,height=3.0cm]{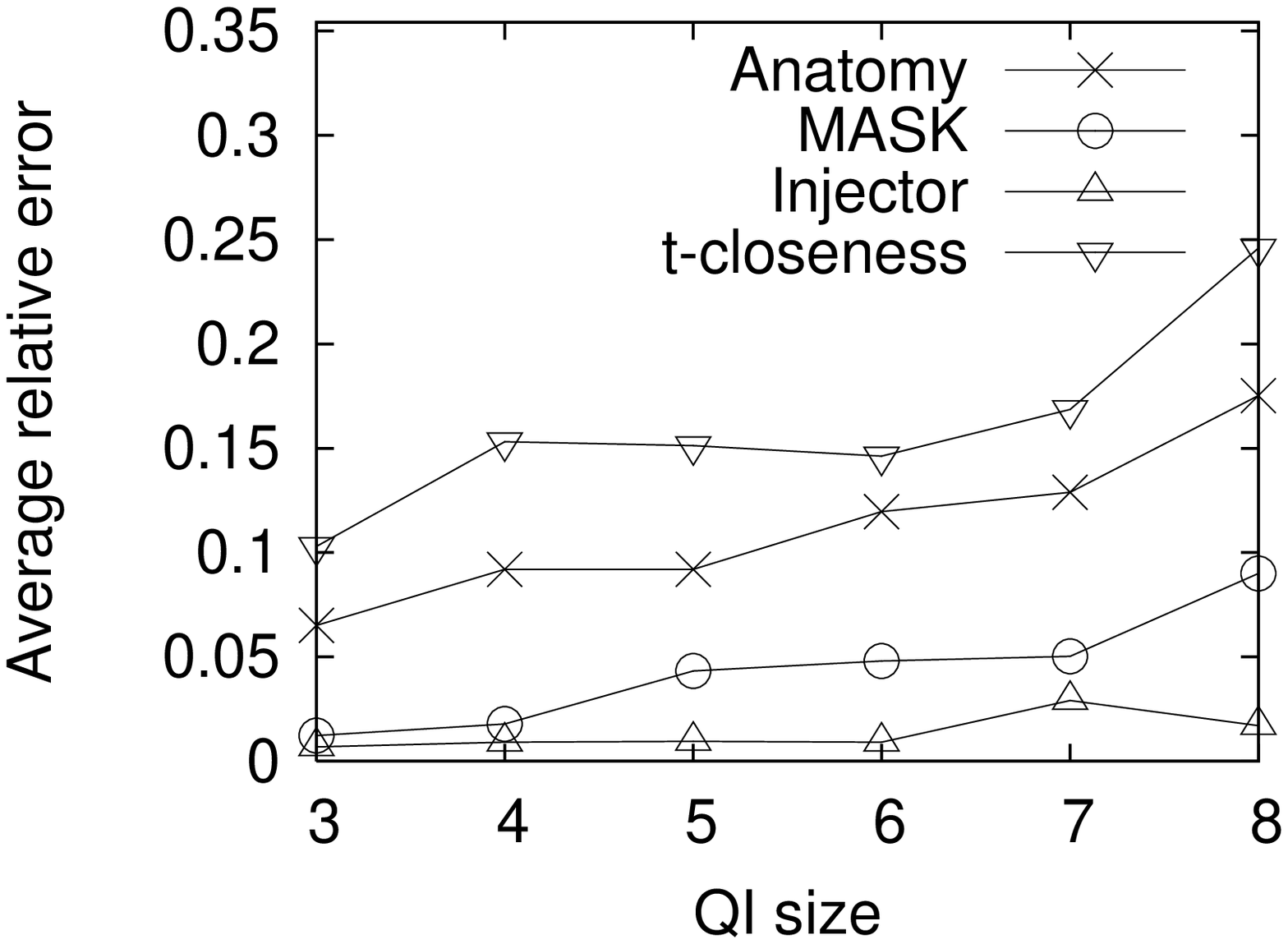}
    \end{minipage}
&
    \begin{minipage}[htbp]{4.0cm}
        \includegraphics[width=4.0cm,height=3.0cm]{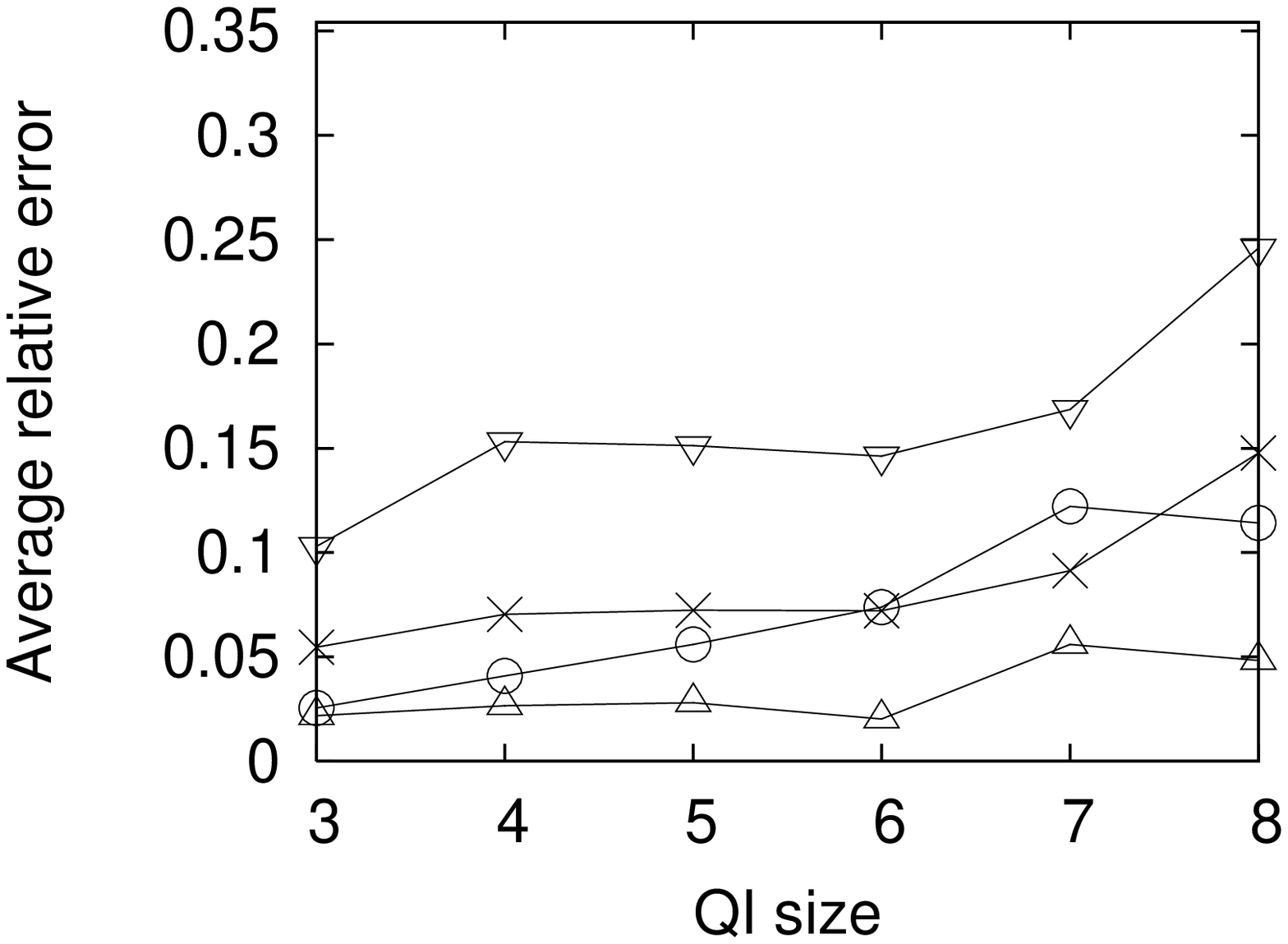}
    \end{minipage}
\\
(a) $r = 2$
&
(b) $r = 10$
\end{tabular}
\caption{Effect of QI size on average relative error}\label{fig:graphAgainstQID-AvgRelativeError}
\end{figure}

We have also conducted experiments when $r = 10$ as shown in
Figure~\ref{fig:graphAgainstQID-m10}. The results are also similar.
But, the time
for mining foreground knowledge is larger. 
Since $r$ is larger and thus $1/r$ is smaller, the
average value of $\triangle$ is smaller
when $r = 10$.
Also, when $r = 10$, there are privacy breaches for
\emph{$t$-closeness} in Figure~\ref{fig:graphAgainstQID-m10}(c)
because there is a higher privacy requirement when $r = 10$ and thus
the size of the A-group is not large enough for protection.

\section{Related Work}
\label{sec:related}

With respect to attribute types considered for
data anonymization, there are two branches of studying. The first
branch is anonymization according to the QI attributes. A typical
model is $k$-anonymity
\cite{AggarwalICDT05,Incognito}. 
The other branch is the consideration of both quasi-identifier
attributes and sensitive attributes. Some examples
are 
\cite{l-diversity}, \cite{WLFW-kdd06}, 
\cite{LL07},
\cite{LL08}
and \cite{BS08}. In this
paper, we focus on this branch. We want to check whether the
probability that each individual is linked to any sensitive value is
at most a given threshold.

%
$l$-diversity \cite{l-diversity} proposes a model where $l$ is a
positive integer and each A-group contains $l$ ``well-represented"
sensitive values. For $t$-closeness \cite{LL07}, the distribution in
each A-group in $T^*$ with respect to the sensitive attribute is
roughly equal to the distribution of the entire table $T^*$. Given a
real number $\alpha \in [0, 1]$ and a positive integer $k$,
$(\alpha, k)$-anonymity  \cite{WLFW-kdd06} maintains that, for each
A-group $L$, the number of tuples in $L$ is at least $k$ and the
frequency (in fraction) of each sensitive value in $L$ is at most
$\alpha$.
%
%

We emphasize that $t$-closeness is different from ours.
Firstly, $t$-closeness does not have any privacy guarantee
on the bound of breach probabilities.
Like $l$-diversity, Anatomy and $(\alpha, k)$-anonymity,
the major goal of privacy protection is to bound
the probability that an individual is linked to a sensitive
value at most a given threshold. However, $t$-closeness
has just an input parameter $t$ expressing the bound on
the closeness between the distribution in each A-group
and the distribution of the entire table, which does not
give any bound of breach probabilities.
Secondly,
$t$-closeness does not consider the QI attribute values
for the distribution. Specifically, the distribution of an A-group
(or the entire table) considered in $t$-closeness
is the global distribution involving the probability that an individual
(with any QI attribute values) is linked to a sensitive value.
However, the global distribution studied here involves
the probability that an individual with particular QI attribute values such as Japanese
is linked to a sensitive value.
Thirdly, enforcing $t$-closeness
gives a large distortion on the anonymized dataset.
This is because it is usually the case that
a small A-group has the distribution
which is very different from the distribution
of the entire table. In order to satisfy $t$-closeness,
a lot of A-groups should be merged to form a very large
A-group, which makes the distortion large.
Fourthly, there are not many useful patterns
found in the table satisfying $t$-closeness.
Like \cite{XT06b,ZKS+07,WFW+07},
one objective to publish the table is to analyze
the correlation between some QI attributes
and the sensitive attribute.
Since $t$-closeness restricts that each A-group
has nearly the same distribution
as the distribution of the entire table,
the desired goal cannot be achieved.

%
%
%

In the literature, different kinds of background knowledge are
considered \cite{l-diversity,KG06,MKM+07,WFW+07,LLZ09,GKS08,LL08,APZ06}.
\cite{KG06} proposes the statistics of some attributes such as
age and zipcode can be also available to the public.
\cite{MKM+07} considers another background knowledge
in form of implications.
\cite{WFW+07} discovers that the minimality principle of the
anonymization algorithm can also be used as a background knowledge.
\cite{LLZ09} proposes to use the kernel estimation method
to mine the background knowledge from the original table.
\cite{GKS08} describes that there are many tables published from
different sources containing overlapping individuals.

\cite{LL08} finds that association rules can be mined from the
\emph{original} table and thus can be used for
privacy protection during anonymization. 
%
In \cite{APZ06}, the problem of privacy attack by adversarial
association rule mining is investigated. Hence, the association
rules
are the foreground knowledge. 
However, as pointed out in \cite{SMB97}, association rules used in
\cite{LL08} and \cite{APZ06} can contradict the true statistical
properties. 
Also the solution in \cite{APZ06} is to invalidate the rules, but
this will violate the data mining objectives of data publication.

A recent work \cite{Aggarwal08} proposes to generate a table in
form of an uncertain data model. 
However, this work 
considers $k$-anonymity which ignores any sensitive attribute.

\section{Conclusion}
\label{sec:concl}

In this paper, we point out a fundamental privacy breach problem
which has been overlooked in the past. With the consideration of the
utility of the anonymized table, group based anonymization suffers
from privacy breaches. Our experiments show that existing well-known
privacy models \emph{Anatomy}, \emph{MASK}, \emph{Injector} and
\emph{$t$-closeness} suffer from serious privacy breaches in a
benchmark dataset.
For future work, we plan to study how to anonymize the data to
defend against foreground knowledge attack. In our experiment, we
observe that the chance of privacy breaches is lower if each group
contains tuples with ``similar" global probabilities. Thus, forming
A-groups with ``similar" tuples is one possible strategy. Another
future work is to
study the effect of background knowledge that may be possessed by
the adversary.

\bibliographystyle{abbrv}
\bibliography{foreground}

\end{sloppy}

\end{document}